# Parametrically amplified phase-incoherent superconductivity in YBa$_2$Cu$_3$O$_{6+x}$

A. von Hoegen[1], M. Fechner[1], M. Först[1], N. Taherian[1], E. Rowe[1], A. Ribak[1], J. Porras[2], B. Keimer[2], M. Michael[3], E. Demler[3], A. Cavalleri[1,4]

[1] *Max Planck Institute for the Structure and Dynamics of Matter, Hamburg, Germany*
[2] *Max Planck Institute for Solid State Research, Stuttgart, Germany*
[3] *Department of Physics, Harvard University, USA*
[4] *Department of Physics, University of Oxford, UK*

**The possibility of enhancing desirable functional properties of complex materials by optical driving is motivating a series of studies of their nonlinear terahertz response. In high-T$_c$ cuprates, large amplitude excitation of certain infrared-active lattice vibrations has been shown to induce transient features in the reflectivity suggestive of non-equilibrium superconductivity. Yet, a microscopic mechanism for these observations is still lacking. Here, we report measurements of time- and scattering-angle-dependent second-harmonic generation in YBa$_2$Cu$_3$O$_{6+x}$, taken under the same excitation conditions that result in superconductor-like terahertz reflectivity. We discover a three-order-of-magnitude amplification of a 2.5-terahertz electronic mode, which is unique because of its symmetry, momentum, and temperature dependence. A theory for parametric three-wave amplification of Josephson plasmons, which are assumed to be well-formed below T$_c$ but overdamped throughout the pseudogap phase, explains all these observations *and* provides a mechanism for non-equilibrium superconductivity. More broadly, our work underscores the role of parametric mode mixing to stabilize fluctuating orders in quantum materials.**

The large amplitude excitation of apical oxygen vibrations in $YBa_2Cu_3O_{6+x}$ has been shown to induce transient optical signatures of superconductivity, evidenced by characteristic edges in time-resolved terahertz reflectivity measurements[1,2,3,4]. Femtosecond soft and hard x-ray scattering measurements[5,6,7] have been applied to clarify the underlying dynamics, and have underscored the role of nonlinearly driven lattice vibrations, the melting of competing charge orders and the transient deformation of the crystal structure. Yet, these findings have not yet provided a comprehensive understanding of this phenomenon.

Here, femtosecond mid-infrared pulses with polarization aligned along the c-axis of $YBa_2Cu_3O_{6.48}$ and $YBa_2Cu_3O_{6.65}$ (see Figure 1a) were used to drive the same 17 and 20 THz apical oxygen oscillations that induce the superconductor-like reflectivity discussed above. Unlike all previous measurements, the carrier-envelope-phase offset of the pump pulses was stabilized[8,9], enabling sub-cycle sampling of the coherent dynamics and yielding both their amplitude and phase. Near-infrared probe pulses of 30-fs duration and polarization aligned with the same *c*-axis direction of the mid-infrared excitation were used to simultaneously record the time-dependent linear reflectivity $\Delta R(t)$ at 800 nm wavelength (Figure 1b) and the time-dependent 400-nm second harmonic intensity $\Delta I_{SH}(t)$ (Figure 1c)[10]. The modulation in $\Delta R(t)$ and $\Delta I_{SH}(t)$ were characterized by a prompt change as pump and probe pulses overlapped in time, followed by a smooth decay and super-imposed coherent oscillations. The reflectivity oscillations $\Delta R(t)$ measured in $YBa_2Cu_3O_{6.48}$ and in $YBa_2Cu_3O_{6.65}$ are reported in the Supplementary Information S1. These have already been discussed in Ref. 11, and reflect the nonlinear excitation of totally-symmetric $A_g$ modes at zero momentum (q=0).[12] Two modes at 3.7 THz and 5 THz are most clearly observed in these measurements.

Time dependent oscillations of the second harmonic intensity $\Delta I_{SH}(t)$, also referred to as *Stimulated Hyperraman Scattering* (SHS), are on the contrary sensitive to coherent symmetry-odd modes[13,14].

SHS is a four-wave-mixing process, mediated by a third-order susceptibility $\chi^3(\omega_{IR}, \omega_{IR}, \omega_{THz})$, which modulates the signal, centered at the optical frequency $2\omega_{IR}$, in time at the frequency $\omega_{THz}$ of the symmetry-odd mode. This is intuitively understood by noting that in a centrosymmetric medium, these modes break inversion as they oscillate and can be equivalently thought of as resulting from a *time-dependent* effective second order susceptibility $\chi^2_{eff}(\omega_{IR}, \omega_{IR}, t)$. It must be stressed that the SHS measurements reported here are fundamentally different from time-integrated second harmonic generation experiments reported extensively in the past[15], which provide signals at the optical frequency $2\omega_{IR}$ through a *time-independent* second-order susceptibility $\chi^2(\omega_{IR}, \omega_{IR})$.

In Figure 2, we report representative SHS data taken in the superconducting state of YBa$_2$Cu$_3$O$_{6.48}$ at T = 5 K (T$_c$ = 48 K). These experiments were conducted for a range of pump electric field strengths between 300 kV/cm (see Figures 2a,b), to 7 MV/cm (see Figures 2e,f). The oscillatory part of the time-resolved second harmonic intensity $\Delta I_{SH}(t)$ was extracted from the traces displayed in Figure 1c by subtracting the slowly varying signal contributions. The data shown in Figure 2a (shaded in yellow), and the corresponding Fourier transform in Figure 2b, are representative of resonant excitation of two modes at 17 THz and 20 THz, which were simultaneously driven by the broad spectrum of the ultrashort pump pulse.

The higher field data reported in Figures 2c,d (500 kV/cm) and 2e,f (7 MV/cm) reflect a nonlinear response regime, where other nonlinearly coupled modes responded to the resonant excitation of the directly driven vibrations. One nonlinearly coupled mode (shaded in red) stands out already at the low fields (Figures 2c,d), oscillating at 2.5 THz frequency, for which no *c*-axis symmetry-odd vibration is expected[16,17,18,19,20]. The amplitude of this mode increased nearly 100-fold as the pump field was increased from 500 kV/cm to 7 MV/cm (Figure 2e,f), evidencing a regime of amplification that will be analyzed in the reminder of the paper.

Other resonances were also observed in the SHS spectrum recorded for the highest excitation fields (see Figure 2e,f), including two infrared active phonons at 8.6 and 10.5 THz (shaded in grey)[19,20] and a broad feature centered around 14 THz, reminiscent of a transverse Josephson Plasma mode[16]. Experiments conducted for $YBa_2Cu_3O_{6.65}$ yielded similar results, with one amplified mode emerging at frequencies where no phonons are expected, oscillating at 2.8 THz as opposed to 2.5 THz for this higher doping level (see Supplementary Information S2).

The multi component oscillations reported in Fig. 2a,c,d were filtered in frequency, and yielded the amplitude of each mode, e.g. $J_1 \propto \Delta I_{SH}(t)_{\omega=2.5\,THz}$ or $Q_{amplified} \propto \Delta I_{SH}(t)_{\omega=8.6\,THz}$. The same could be done for $Q_{drive} \propto \Delta I_{SH}(t)_{\omega=17\,THz}$ and we could then extract the growth of the amplified modes as a function of the drive $J_1(Q_{drive})$ and $Q_{amplified}(Q_{drive})$, as displayed in Figure 2g. We found that above a characteristic threshold value for the drive amplitude, each nonlinearly coupled oscillation grew exponentially, and was amplified by at least three orders of magnitude in amplitude (see Fig. 2g).

In Figure 2h,i we report the temperature ($T$) dependence of $J_1(Q_{drive}, T)$ at the two doping levels analyzed here and $Q_{amplified}(Q_{drive})$ (see corresponding time-domain oscillations and Fourier transforms in Supplementary Information S3). These data show that the 2.5 THz oscillations in $YBa_2Cu_3O_{6.48}$ and the 2.8 THz mode in $YBa_2Cu_3O_{6.65}$ do not follow the temperature dependence of any one of the equilibrium modes. Indeed, in equilibrium one finds a 1 THz frequency Josephson Plasma Resonance, which disappears at $T_c$ (see dashed curve in Figure 2h), and a number of phonon resonances, which remain large above $T_c$ (see dashed curve in Figure 2i). On the other hand, the quantity $J_1(T)$ plotted in Figure 2h decayed only at a characteristic temperature scale T'. An empirical fit function proportional to $\sqrt{1-T/T'}$ yielded $T' \approx 380$ K and $T' \approx 280$ K for $YBa_2Cu_3O_{6.48}$ and for the $YBa_2Cu_3O_{6.65}$, respectively. These temperature scales agree well with the corresponding pseudogap temperatures $T^*$ for each doping[21]. Finally, this temperature dependence is different from that of the amplified infrared active phonons (see Figure 2i), which follow the temperature dependence of the corresponding equilibrium phonon modes[20], remaining of constant amplitude for all temperatures and exhibiting only an anomaly at $T_c$.

The symmetry of the 2.5 THz mode was measured as sketched in Figure 3a, by repeating all the SHS measurements of Figure 2 at a number of polarization angles $\varphi$ for the incoming 800-nm probe pulses, whilst the 400-nm second harmonic intensity was measured with an analyzer along the c-axis (s polarization). All the data reported in Figure 3 were acquired at room temperature, where the results were identical to those collected at $T < T_c$. Corresponding measurements with the analyzer aligned along the b axis (p polarization) are reported in the Supplementary Information S4. These SHS polarimetry measurements yielded two-dimensional maps as a function of polarization angle and pump probe time delay, as shown in the representative plot of Figure 3b. With the knowledge of the different mode frequencies extracted from the experiments of Figure 2, we filtered these two-dimensional plots in frequency, and extracted time-resolved SHS polarimetry measurements for each mode. In Figures 3c-e we compare the symmetry of the driven 17 and 20 THz phonons (yellow), of the amplified 8.6 and 10.5 THz phonons (grey) and of the 2.5 THz mode (red), displayed at one representative time delay ($t = 150$ fs).

As shown in Figures 3c,d, the directly driven phonons (yellow) and the amplified phonons (grey) reflect a *Pmm2* space group, that is one in which one mirror symmetry of the equilibrium *Pmmm* space group is broken periodically during the oscillation. This is consistent with the behaviour expected for a $B_{1u}$-symmetry distortion of these phonons. In contrast with all driven and amplified phonons, the angular dependence of the 2.5 THz mode shown in Figure 3e, is unique, being indicative of a lower symmetry for which at least two mirrors of the equilibrium $YBa_2Cu_3O_{6+x}$ structure are lost (*Pm* or lower symmetry). In addition, and in stark contrast to the phonons, the angular dependent response exhibits a number of nodes, with the electronic mode changing sign as a function of polarization angle $\varphi$.

We further investigated the nature of the room temperature 2.5 THz mode by measuring SHS spectra as a function of outgoing scattering angle, a reporter of their in-plane momentum. SHS measurements are sensitive to coherent oscillations at far shorter wavelengths than those accessible by linear spectroscopy. The 400 nm second harmonic emission is for example sensitive to momenta in excess of $10^4$ cm$^{-1}$, which for a 2.5 THz excitation is far larger than the corresponding light cone momentum accessible by linear optics (~ 80 cm$^{-1}$). As sketched in Figure 4a, these momentum-resolved measurements were obtained by aligning the mid-infrared pump and the 800-nm probe pulses collinearly with one another, and by measuring the SHS signal for a series of outgoing scattering angles using a slit (see Supplementary Information S5). The crucial observation is that whilst all directly driven (yellow) and amplified (grey) phonon signals peaked for specular reflection, corresponding to driven or amplified vibrations with zero in-plane momentum ($q_y = 0$), the 2.5 THz mode (red) peaked at finite in-plane momenta $q_y = \pm 190$ cm$^{-1}$.

In the following, we present a theory that connects the observed amplified 2.5 THz (2.8 THz) modes to the nonlinear physics of Josephson Plasmon Polaritons (JPP)[22], dispersive c-axis modes made up of interlayer tunnelling currents and propagating along the CuO$_2$ planes[23]. In bilayer YBa$_2$Cu$_3$O$_{6+x}$, two JPP modes are found below T$_c$[24,25]. At zero momentum, these two modes are dominated by current flow between bilayers (lower frequency mode) and within the bilayers (higher frequency mode), depicted in Figure 5d. The zero-momentum frequency for these modes are well known, although their momentum dependence has never been documented. As the momentum along the propagation direction changes, these JPP modes are no longer solely made up of c-axis currents, but involve also in-plane superflow, with a dispersion determined by the inductive response of the planes. In Figure 5a,b we display their computed equilibrium dispersion curves, obtained from known zero momentum frequencies and in-plane inductance (see Supplementary Information S7).

To derive the nonlinear equations of motion for these JPP modes, and the coupling to the driven lattice vibrations, one first needs to supplement Maxwell equations with the relations between the supercurrents and electromagnetic fields[26]. The latter can be obtained directly from the kinetic energies of the interlayer tunnelling and of the in-plane superflow, referred to here as $E_{Jtunn}$ and $E_{Jplane}$, respectively. The expression for these energies is $E_{Jtunn} = -J_{n,n+1}\cos\left[\theta_n - \theta_{n+1} - \frac{2e}{c}\int_{z_n}^{z_{n+1}} A_z dz\right]$ and $E_{Jplane} = \frac{1}{2}\rho_s v_s^2$. Here, $\theta_n(x,y,t)$ is the order parameter within each $CuO_2$ layer $n$, $\rho_s$ denotes the local superfluid density and $v_s = \nabla_{x,y}\theta_n - \frac{2e}{c}A$ is the in-plane superfluid velocity, which itself is a function of the in-plane order parameter gradient $\nabla_{x,y}\theta_n$ and of the vector potential $A$. In these expressions, $2e$ is the Cooper pair charge and $c$ the speed of light.

The apical oxygen phonons excited by the pump to amplitude $Q_{drive}$ are symmetry odd, and therefore modify the local superfluid densities in the $YBa_2Cu_3O_{6+x}$ bilayer structure in a way that is antisymmetric with respect to the two layers, $\delta\rho_{s\{1,2\}} \propto \pm Q_{drive}$[26]. The effect of these vibrations on the in-plane superflow is then to increase and decrease the in plane kinetic energy $E_{Jplane}$ in neighbouring planes in an oscillatory fashion, which can be written as $\delta E_{Jplane} = \delta\rho_{s1}v_{s1}^2 + \delta\rho_{s2}v_{s2}^2$ (see Supplementary Information S7 for details). This expression can be rewritten as $\delta E_{Jplane}(t) \propto Q_{drive}(t)(v_{s1} - v_{s2})(v_{s1} + v_{s2})$ or, equivalently, as $\delta E_{Jplane}(t) \propto Q_{drive}(t)J_1J_2$, as the apical oxygen oscillations of coordinate $Q_{drive}$ and the two finite-momentum tunnelling modes are excited with $J_1 \propto v_{s1} - v_{s2}$ and $J_2 \propto v_{s1} + v_{s2}$, respectively. Intuitively, the c-axis currents $J_1$ and $J_2$ are driven by the lattice excitation because the changes in the in-plane kinetic energy also perturb the in-plane gradients of the order parameter phase $\nabla_{x,y}\theta_n$, which then, through the second Josephson relation $J_{n,n+1} = J_c sin(\Delta\theta_{n,n+1})$, makes the c-axis tunnelling dependent on the in-plane spatial coordinate. The equations of motion for the JPPs are then

$$\ddot{J}_1 + 2\gamma_{J_1}\dot{J}_1 + \omega_{J_1}^2(q_{x1}, q_{y1})J_1 = -aq^2 Q_{drive}(t)J_2$$

$$\ddot{J}_2 + 2\gamma_{J_2}\dot{J}_2 + \omega_{J_2}^2(q_{x2}, q_{y2})J_2 = -aq^2 Q_{drive}(t)J_1$$

where $\omega_{J_i}(q_{xi}, q_{yi})$ describe the in-plane equilibrium dispersion. These equations predict three-wave mixing between the apical oxygen phonons $Q_{drive}$ and the upper and lower JPPs, leading to the excitation of damped harmonic oscillations for $J_1$ and $J_2$ at finite momenta along the two-dimensional dispersion curves of Figure 5a, with a driving term $aq^2 Q_{drive}(t)J_{2,1}$. Note that the driving term depends on the momentum of the JPP as $q^2$, hence it vanishes for long wavelengths ($q = 0$) but naturally couples to supercurrents at finite in-plane wavevectors, as observed experimentally in the measurements of Figure 4. The two equations predict further that the phonon excites pairs of JPPs with frequencies that satisfy $\omega_1 + \omega_2 = \omega_{drive}$, and opposite in plane momenta ($q_{x1} = -q_{x2}$ or $q_{y1} = -q_{y2}$). A numerical solution of these equations of motion is displayed in the color-coded dispersion of Figure 5a,b. There, the three-wave mixing process is shown to couple the driven phonon to JPPs at in-plane momenta $q_y \sim 200$ cm$^{-1}$ (see Figure 5c) where phase matching is fulfilled. Note that the momentum at which excitation of $J_1$ and $J_2$ is expected matches very well the experimental findings reported in Figure 4, despite it being obtained here from first principles and starting from the documented inductive response of $YBa_2Cu_3O_{6+x}$. Since the mid-infrared pump field and hence the driven phonons extend in the direction perpendicular to the optical surface only over a skin depth of ~1.5 µm, phase matching is inefficient along x. Hence, the excited pairs of high (intra bilayer) and low (inter bilayer) frequency JPPs propagate along the optical surface like $J_1(\omega_1, +q_y)$ and $J_2(\omega_2, -q_y)$, or $J_1(\omega_1, -q_y)$ and $J_2(\omega_2, +q_y)$.

The theoretically predicted energy and momentum matching is consistent with the experimentally determined frequency resonance ($\omega_{drive} \sim 17$ THz, $\omega_1 \sim 2.5$ THz, $\omega_2 \sim 14.5$ THz), documented in Supplementary Information S6, and is also consistent with the reported momentum resolved measurements of Figure 4. Furthermore, the calculated exponential amplification of the JPP, shown in Figure 5e, resembles the experimental results reported in Figure 2. Finally, the calculated real-space current patterns of these finite-momentum modes are sketched in Figure 5f and compared to the real-space distortions of the driven apical oxygen mode in Figure 5g. As explained by the theory, in-plane currents (drawn as black arrows) drive finite momentum c-axis tunnelling currents $J_1$ and $J_2$ (dashed lines), which break the two mirror-planes perpendicular to the CuO$_2$ layers (m$_{z,y}$ and m$_{z,x}$), in agreement with the SHS polarimetry signal reported in Figure 3.

In the following, we show that this theory naturally explains also the transient reflectivity edges reported in previous time resolved terahertz probe experiments. The three-wave mixing excitation results in two counterpropagating Josephson Plasma Polaritons with opposite momentum $\pm q_{JPP}$, which interfere to produce a standing wave pattern of the superconducting phase $\theta$ along the sample surface $y$ direction (see Fig. 6a) as $\theta(y,t) = \theta_0 \cos(q_{JPP} \cdot y) \sin(\omega_{JP} \cdot t)$. Through the Josephson equations, the dynamics of the phase $\theta(y,t)$ can be re-casted as a source of modulation of the in-plane superfluid density $\rho_s(y,t)$ at q$_y$ = 0 as $\rho_s(y,t) = \rho_{s,0} \cdot \cos(\theta(y,t))$, sketched in Figure 6b. For finite excursions of the condensate phase ($\theta(y,t) \ll \pi/2$) two salient features emerge at lowest order expansion, that is a spatial modulation of the superfluid density at $2q_{JPP}$ and a temporal modulation at $2\omega_{JPP}$ for in plane momentum $q_y = 0$

$$\rho_s(y,t) \approx \rho_{s,0}[1 - \{\theta_0^2 + \theta_0^2 \cos(2q_{JPP} \cdot y) - \theta_0^2 \cos(2\omega_{JPP} \cdot t) - \theta_0^2 \cos(2q_{JPP} \cdot y)\cos(2\omega_{JPP} \cdot t)\}/4].$$

As sketched in Figure 6b, this zero-momentum modulation of the superfluid density $\rho_s$ at $2\omega_{JPP}$ naturally leads to parametric amplification of zero-momentum plasma waves at $\omega_{JPP}$, and by extension of an optical probe at the same frequency. Note that the parametric amplification of Josephson plasma waves was experimentally demonstrated in a related setting in Ref. 27 and is taken here as being a reliable conjecture.

From these equations, the transient THz reflectivity[1,2,3,4] can be computed both below and above $T_c$. Below $T_c$, as shown in Figure 6d and 6e, the theory predicts an additional plasma edge at a frequency $\sim 2\omega_{JPP} \sim 2$ THz, above the equilibrium Josephson plasma edge at $\sim 1$ THz. This prediction compares well with the experimental reports, summarized in Figures 6h,i. In the temperature range $T_c<T<T^*$, the featureless equilibrium THz reflectivity is retrieved in these simulations by assuming the same interlayer Josephson coupling strength observed below $T_c$ with a sizeable increase in de-phasing, as to make the Josephson plasmon overdamped (see Figure 6e)[26]. For a strong drive of the apical oxygen phonon, the plasma edge is predicted to re-emerge at $\omega_{JPP} \sim 2$ THz (see Figures 6e,f), again blue shifted with respect to the below-$T_c$ equilibrium resonance. At this frequency parametric driving compensates dissipation most efficiently and revives the features of the dissipation-less state. In Figures 6j,k we compare these calculations to experimental THz reflectivity data reported for a $YBa_2Cu_3O_{6.5}$ sample in Ref. 2 and find good agreement.

We conclude by noting that other trends in the data are supported by the line of argument laid out in this paper. The observed doping dependence for the induced reflectivity edge reported in Refs. 2 and 4 involves a response that extends to T* for underdoped regions of the spectrum, but becomes negligible at higher doping, falling below T*. The observed trends follow naturally from the requirement of a pre-existing phase incoherent superconductivity *and* the need of resonant three-wave mixing discussed in Figure 5. Because the three-wave mixing mechanism requires that

$\omega_{drive} = \omega_{J1} + \omega_{J2}$, with the in-plane momentum chosen to fulfill this condition along the dispersing branches of the JPPs, this effect is only possible for doping levels for which $\omega_{J1,q=0} + \omega_{J2,q=0} < \omega_{drive}$. This condition is fulfilled for YBa$_2$Cu$_3$O$_{6.48}$ and YBa$_2$Cu$_3$O$_{6.65}$, with $\omega_{J1}^{6.48}$ = 2.5 THz and $\omega_{J1}^{6.65}$ = 2.8 THz, but not for higher doping levels, where $\omega_{J1} + \omega_{J2} > \omega_{drive}$ at all momenta. Indeed, results in YBa$_2$Cu$_3$O$_{6.92}$ reported in the Supplementary Information S2 confirm that no amplification of Josephson Plasmon Polaritons takes place at these doping values, also in agreement with the lack of a light-induced reflectivity edge reported in Refs. 2 and 4, even for those doping values in which T* > T$_c$.

The body of evidence and theoretical analysis discussed above also provides useful perspective for the physics of high-T$_c$ cuprates in general. Whilst the present measurement is not comprehensively addressing the entire pseudogap region of the phase diagram, in the doping range where the parametric resonance between phonons and Josephson Plasmons can be fulfilled, our results can convincingly be explained if one assumes pre-existing fluctuating superconductivity up to T*, over correlation lengths of several microns[28]. As support for this hypothesis one should mention that although optical signatures of superconductivity at large correlation lengths vanish at T$_c$[29,30], momentum integrated probes like the superconducting Nernst effect[31,32,33,34,35], measurements of electrical noise[36] and optical spectroscopy[37], provide evidence of residual coherence in the normal state, to be validated by more comprehensive methods such as high-resolution Resonant Inelastic X-ray Scattering[38,39] and Electron Energy Loss probes[40]. More broadly, our experiments open up new perspectives of frequency resonant wave mixing as a new means to control cooperative phenomena in quantum materials.

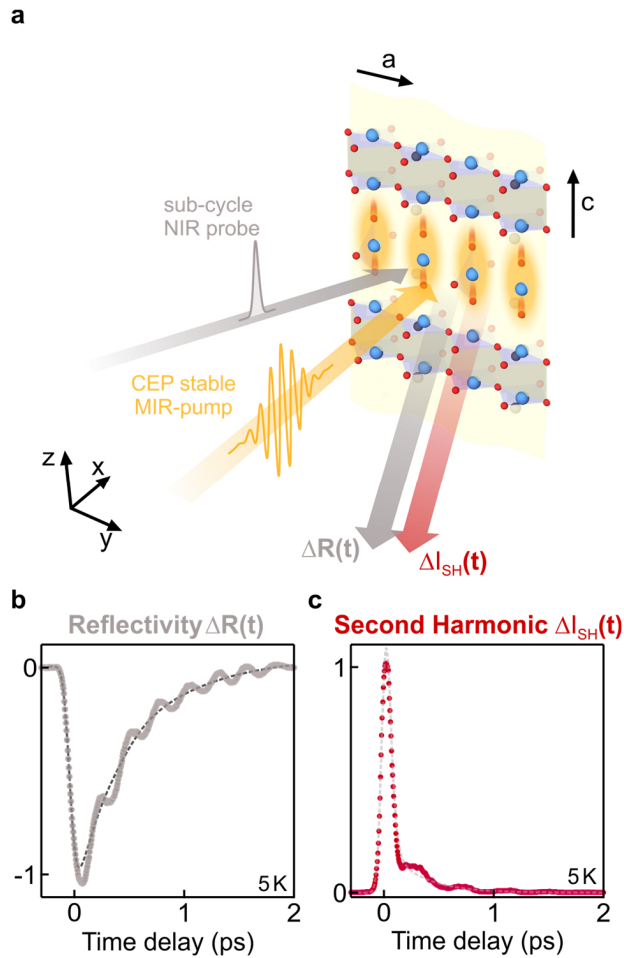

**Figure 1| Optical probe-pump geometry and time-resolved optical changes. a,** Schematic of the probe geometry. The 800-nm near-IR probe (grey) and CEP stable mid-IR pump (yellow) pulses were polarized along the $YBa_2Cu_3O_{6.48}$ crystals c-axis and perpendicular to the $CuO_2$-planes. The resonantly excited apical oxygen vibrations are shaded in yellow. Light at the fundamental (grey arrow) and second harmonic (red arrow) frequency is reflected from the sample. **b,** Time-resolved polarization rotation of the linear reflectivity at 800 nm wavelength, showing coherent modulations due to fully symmetric Raman phonon modes. **c,** Time resolved second harmonic intensity at 400 nm wavelength (red circles) and a numerical fit to the non-oscillatory component of the signal (dashed line). The data in b and c are shown for 5 K base temperature.

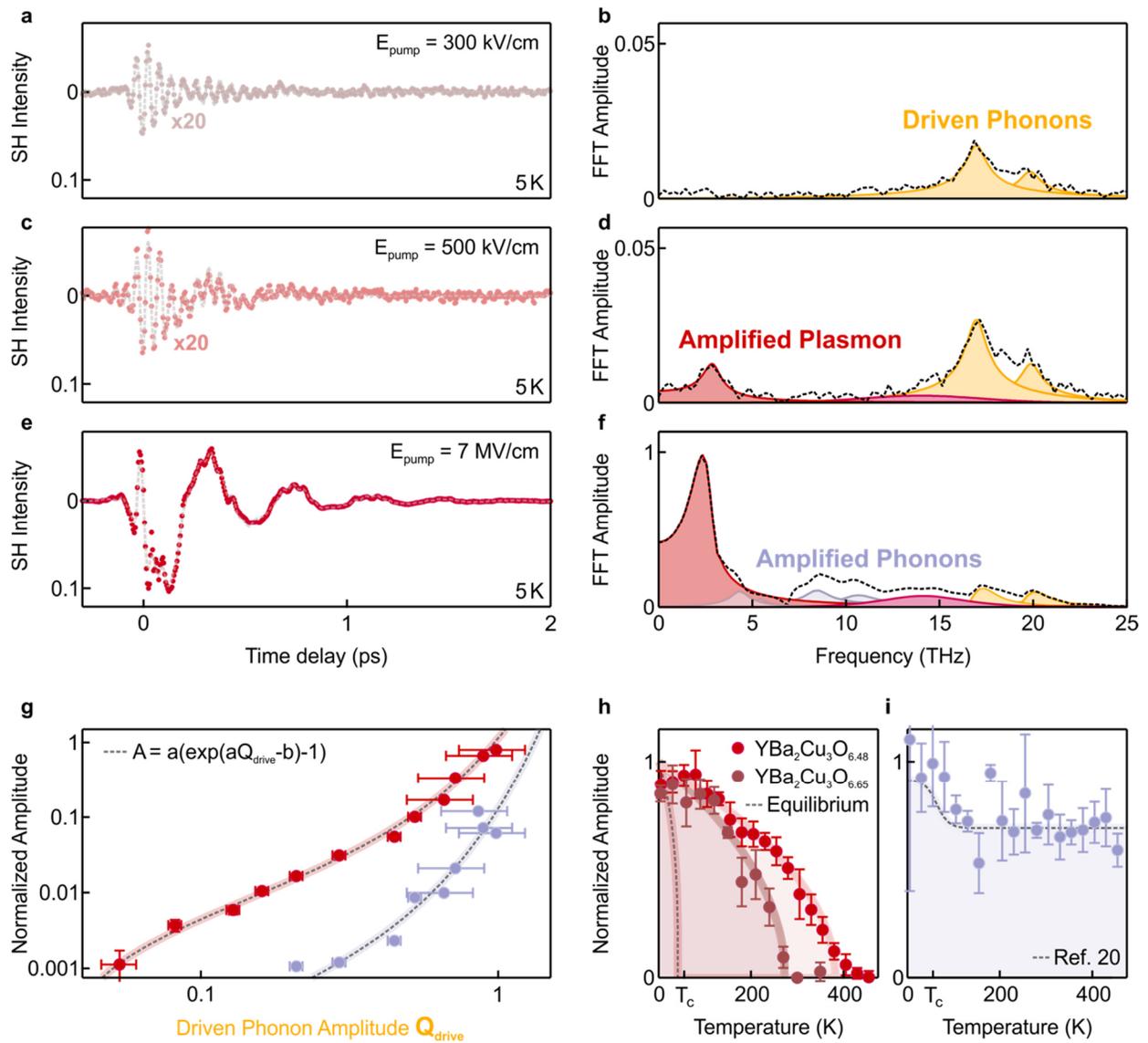

**Figure 2| Coherent oscillations in the time-resolved second harmonic intensity. a** and **b**, Coherent SHS signal at the lowest excitation field (E = 0.3 MV/cm) and the corresponding Fourier amplitude spectrum at T = 5 K, well below the critical temperature $T_C$ = 48 K. The high frequency oscillations at 17 and 19.5 THz (yellow peaks) are coherent symmetry breaking apical oxygen vibrations, resonantly driven by the excitation pulse. **c** and **d**, Coherent SHS response at higher excitations fields (E = 0.5 MV/cm) at the same temperature. The peaks at $\nu_1$ = 2.5 THz and $\nu_2$ = 14 THz (red and magenta) are ascribed to coherent oscillations of Josephson Plasma modes. **e** and **f**,

The coherent SHS response at significantly stronger excitations (E = 7 MV/cm) show the same coherences of panel c and b, with additional modes drawn as grey peaks. These additional peaks are dominated by those at 8.6 and 10.5 THz and label additional phonons nonlinearly coupled to the resonantly driven lattice modes. **g**, Measured amplitude of the amplified low frequency Josephson Plasmon $J_P$ and the amplified phonon amplitude $Q_{amplified}$ plotted as a function of the driven apical oxygen vibration amplitude $Q_{drive}$. All quantities were extracted from the same time-trace, for different strengths of the mid-infrared excitation field. The dashed line is an exponential fit $A(Q) = a \cdot (e^{\alpha Q_{drive} - \beta} - 1)$ to the data. Error bars represent the standard deviation $\sigma$ of the amplitudes extracted by numerical fits. **h**, Full temperature dependence of the Josephson plasmon peaks in panel d (shaded area) for $YBa_2Cu_3O_{6.48}$ (red) and $YBa_2Cu_3O_{6.65}$ (dark red). The lines are fits to the data with a mean-field approach $\propto \sqrt{1 - T/T'}$, yielding $T' = 380$ K for $YBa_2Cu_3O_{6.48}$ and $T' = 280$ K for $YBa_2Cu_3O_{6.65}$. The dashed line is the temperature dependence of the equilibrium low-frequency Josephson plasma resonance in $YBa_2Cu_3O_{6.5}$, which disappears at $T_c$. **i**, Temperature dependence of the amplitude of the nonlinearly coupled infrared active phonons at 8.5 and 10 THz. Their temperature dependence from equilibrium infrared measurements in $YBa_2Cu_3O_{6.5}$, taken from Ref. 20, is shown as a dashed line. Error bars represent the standard deviation $\sigma$ of the amplitudes obtained by repeating the experiment under equivalent conditions.

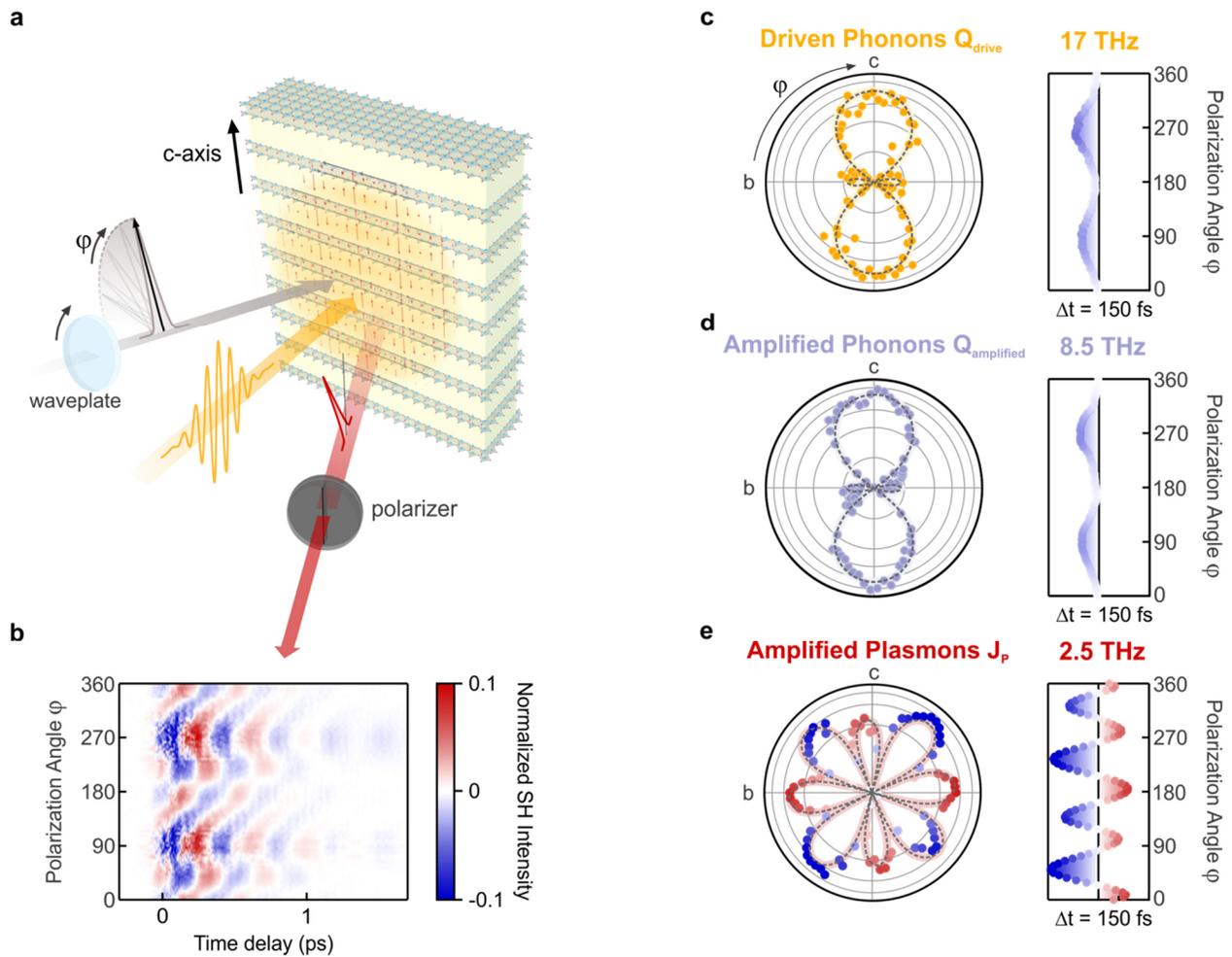

**Figure 3| SHS polarimetry of the coherent oscillations. a,** Schematic of the SHS polarimetry geometry. The polarization angle $\varphi$ of 800-nm NIR probe pulses (grey) is controlled by rotating a $\lambda/2$-waveplate. The reflected second harmonic light (red arrow) passes through an analyzer to measure the different polarization components individually. **b,** SHS signal as a function of polarization angle and pump-probe time delay, measured at room temperature. **c,d** and **e,** Normalized polarimetry signal of the driven phonons (yellow dots), amplified phonons (grey dots) and amplified Josephson plasma modes (red and blue dots) for an analyzer oriented along the

crystals c-axis, at one time-delay t = 150 fs. The SHS polarimetry signal of the two sets of phonons can be reproduced by a fit to a second harmonic tensor with *mm2* point group symmetry (dashed line) and the phase of the oscillations is polarization angle $\varphi$ independent. The polarimetry signal of the amplified Josephson plasmon agrees with a fit to a point group of lower symmetry (dashed line), in which at least two mirrors are lost. The phase of the coherent oscillations is periodically modulated with polarization angle $\varphi$, indicated by the red and blue color-coding.

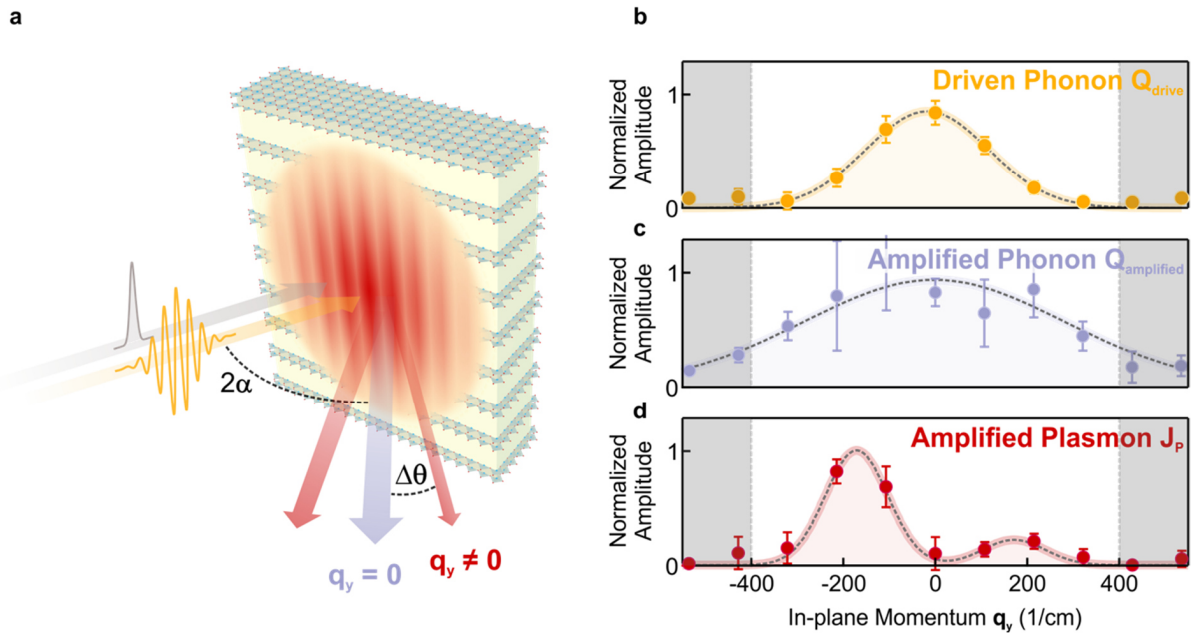

**Figure 4| Spatial dependence of the SH emission. a**, Schematic of the collinear geometry, where pump and probe beams are aligned collinear to excite the sample at an incidence angle $\alpha$ to the surface normal. The specular linear reflection at 800 nm (grey arrow) obeys Snell's law and leaves the sample with the same angle $\alpha$. A finite momentum transfer, due to scattering of propagating modes, appears as a deflection $\Delta\theta$ from the specular deflection (red arrows). **b,c** and **d**, Momentum distribution of the driven phonons (yellow dots), amplified phonons (grey dots) and amplified Josephson plasma modes (red dots), measured at room temperature. The shaded grey areas denote the maximum accessible momentum of the experiment. Error bars represent the standard deviation $\sigma$ of the amplitudes extracted by numerical fits.

# Theory

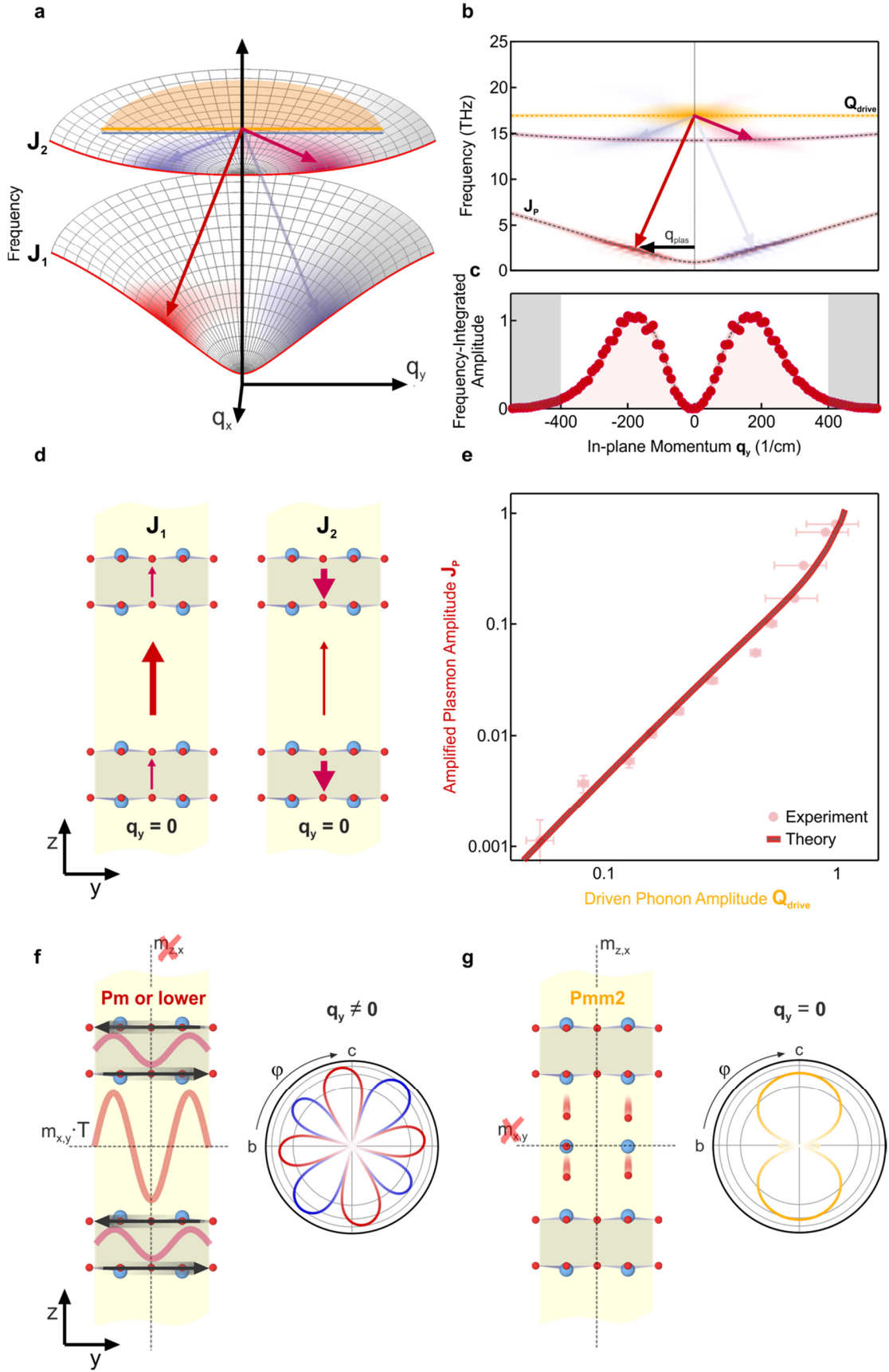

**Figure 5| Theoretical model of nonlinear phonon-Josephson plasmons coupling. a**, Dispersion of the inter ($J_1$) and intra-bilayer ($J_2$) Josephson plasma modes along the in-plane momenta $q_x$ and $q_y$ in $YBa_2Cu_3O_{6.5}$. The red lines are a cut through the $q_x = 0$ plane. The apical oxygen phonon mode at 17 THz (yellow) does not disperse along either direction. The three-wave scattering process is sketched as red and blue arrows and results from a numerical simulation in response to the resonant drive of the apical oxygen phonon at $q = 0$ are shaded in the same colors. The response vanishes along $q_x$, parallel to the light propagation direction. **b**, Detailed insight into the simulation results along $q_y$ for $q_x = 0$. The driven phonon with zero momentum excites a pair of Josephson Plasmon Polaritons, $J_1$ and $J_2$, with opposite wavevectors $q_y$ and frequencies that add up to the phonon frequency. The two processes for mirrored momentum transfer are shown as red and blue arrows, respectively. **c**, Amplitude of the Josephson plasma modes obtained integrating along the vertical frequency axis of panel b. The amplitude is zero for $q_y = 0$, and peaks at $q_y = \pm 200$ cm$^{-1}$. **d**, Sketch of the two Josephson Plasma modes at $q = 0$, with the supercurrents oscillating in-phase ($J_1$) or out-of-phase ($J_2$) for the low and high frequency mode, respectively. The thicknesses of the arrows indicate the supercurrent strengths within and between the bilayers. **e**, Simulated excitation-strength dependence (dashed line) of the low frequency 2.5 THz oscillations, together with the experimental data (dots) from Figure 2g. **f** and **g**, Sketch of the real space symmetries of the finite momentum amplified Josephson plasma modes (*m* or lower) and the driven apical oxygen vibration (*mm2*). The in-plane currents (black arrows) and the resulting finite-momentum Josephson plasma modes (red and magenta lines) break the x,z and y,z mirror-planes. The optical $B_{1u}$-symmetry phonons break the x,y mirror plane only.

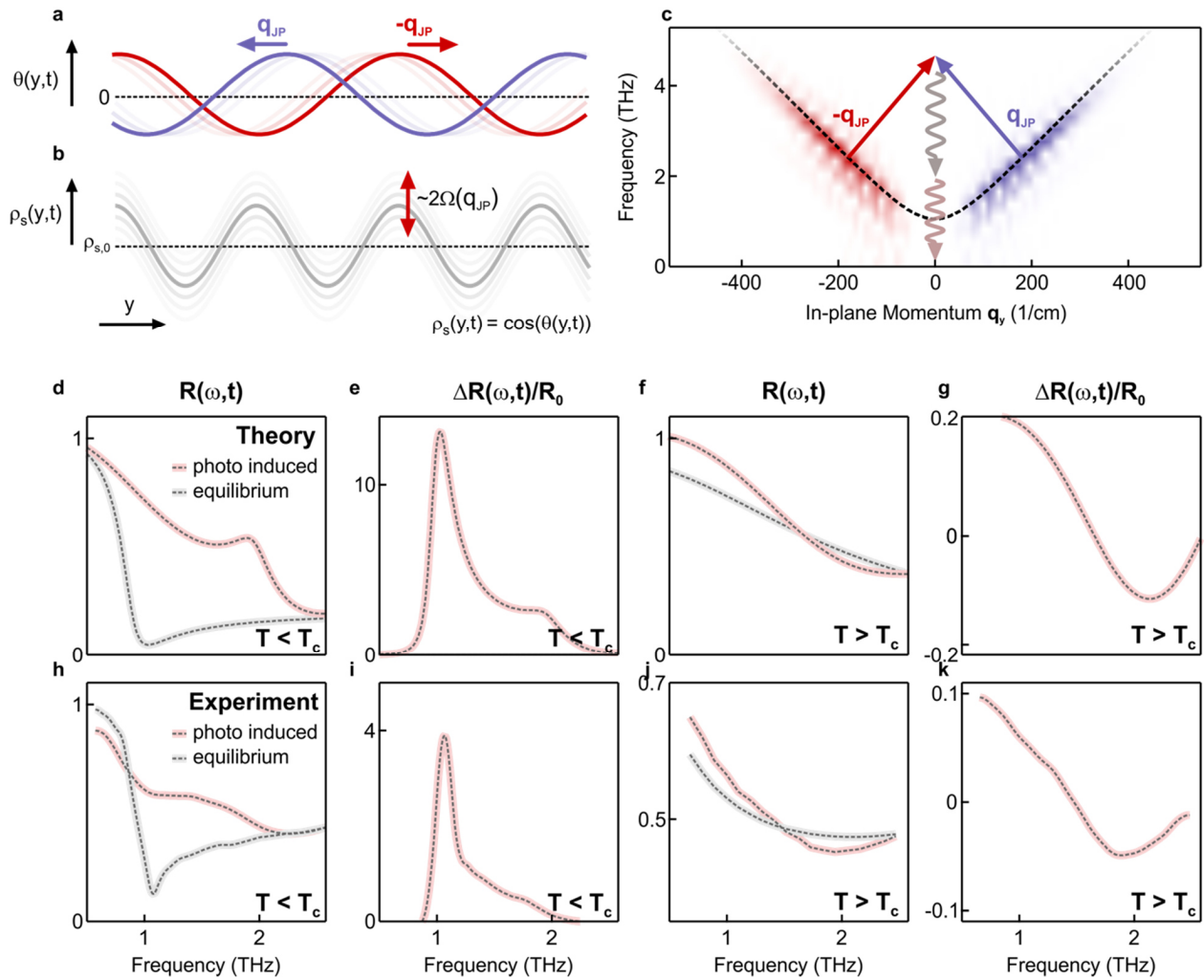

**Figure 6| Floquet-THz-optical properties of the light induced finite-momentum Josephson plasma. a**, Two counter-propagating Josephson plasma modes with in-plane momentum $q_{JP}$ periodically modulate the superconcucting order parameter $\theta(y,t)$ along the y-axis. **b**, Resulting spatial and temporal (shaded grey) modulation of the superfluid density $\rho_s(y,t)$ around its equilibrium value $\rho_{s,0}$. The zero-momentum temporal modulation proceeds at twice the Josephson plasma frequency $2\Omega(q_{JP})$. **c**, Frequency-momentum diagram of the parametric amplification that drives the THz emission. The two counter-propagating plasmons (red and blue arrows) enter into a virtual zero-momentum state, from which two THz photons are emitted. **d** and **e**, Calculated photo-induced (red) and equilibrium (grey) THz reflectivity below $T_c$ and the normalized

reflectivity changes $\Delta R(\omega,t)/R_0$, respectively. The finite-momentum Josephson Plasmon Polariton induces a second plasma edge at a frequency close to $2\Omega(q_{JP})$ and overall enhancement of the reflectivity above the equilibrium plasma edge. **f** and **g**, Calculated photo-induced (red) and equilibrium (grey) THz reflectivity above $T_c$ and the normalized reflectivity changes $\Delta R(\omega,t)/R_0$, respectively. Excitation of the apical oxygen vibration leads to the appearance of a plasma edge at a frequency close to $2\Omega(q_{JP})$, which is clearly visible in the normalized reflectivity changes $\Delta R(\omega,t)/R_0$. **h**, **i**, **j** and **k**, show the same features, observed experimentally in time-resolved THz-reflectivity measurements after resonant excitation of apical oxygen oscillations (Ref. 2).

# Supplementary Materials for

# Parametrically amplified phase-incoherent superconductivity in YBa$_2$Cu$_3$O$_{6+x}$

A. von Hoegen[1], M. Fechner[1], M. Först[1], N. Taherian[1], E. Rowe[1], A. Ribak[1], J. Porras[2], B. Keimer[2], M. Michael[3], E. Demler[3], A. Cavalleri[1,4]

[1] *Max Planck Institute for the Structure and Dynamics of Matter, Hamburg, Germany*
[2] *Max Planck Institute for Solid State Research, Stuttgart, Germany*
[3] *Department of Physics, Harvard University, USA*
[4] *Department of Physics, University of Oxford, UK*

**S1. Nonlinear phonon-phonon coupling**

The phonon spectrum of the ortho-II structure of YBa$_2$Cu$_3$O$_{6.5}$ consists of 73 non-translational modes at the Brillouin zone center. The most relevant phonon modes for *c*-axis polarized THz and mid-infrared excitation are 13 infrared-active B$_{1u}$ modes and 11 Raman-active A$_g$ modes. The full lattice potential $V_{lattice}$ consists of three distinct contributions.[5,11,41,42]

1. The harmonic potential of each phonon mode

$$V_{harm} = \Sigma \frac{\omega_i^2}{2} Q_i^2,$$

    with $\omega_i$ and $Q_i$ representing the eigenfrequency and coordinate of the $i$-th mode, respectively.

2. The anharmonic potential containing higher-order terms of the phonon coordinates and combinations of different phonon modes

$$V_{anharm} = \Sigma g_{ijk} Q_i Q_j Q_k + \Sigma f_{iklm} Q_i Q_k Q_l Q_m,$$

    with $g_{ijk}$ and $f_{iklm}$ being the third-order and fourth-order anharmonic coefficients.

3. The coupling of the resonantly driven infrared-active $B_{1u}$ phonon mode (with coordinate $Q_{drive}$) to an external electric field $E_{field}$

$$V_{efield} = \sum Z^*_{drive} Q_{drive} E_{field},$$

with $Z^*_{drive}$ representing the mode effective charge.

The crystal lattice dynamics are then determined by the equations of motion for each phonon mode, given by

$$\ddot{Q}_i + 2\gamma_i \dot{Q}_i + \nabla_{Q_i}(V_{harm} + V_{anharm} + V_{efield}) = 0.$$

Here $\gamma_i$ is a phenomenological damping term, which accounts for contributions to the finite lifetime not already considered within the anharmonic potential. The equations are restricted to phonon modes at the Brillouin zone center, due to the negligible momentum of long-wavelength THz light.

First, we consider the impact of the third-order terms in $V_{anharm}$. Only the 11 $A_g$ modes fulfill the symmetry requirements to exhibit third-order coupling to the driven $B_{1u}$ mode.[5,11] To simplify the discussion, we consider coupling between the driven mode and a single $A_g$ mode. The equations of motions then reduce to

$$\ddot{Q}_{drive} + \gamma \dot{Q}_{drive} + \omega^2_{drive} Q_{drive} + 2g Q_{drive} Q_{Ag} = Z^*_{drive} Q_{drive} E_{field}$$

$$\ddot{Q}_{Ag} + \gamma_{Ag} \dot{Q}_{Ag} + \omega^2_{Ag} Q_{Ag} + g Q^2_{drive} = 0.$$

They describe a process known as ionic Raman scattering which entails a transient displacement and superimposed oscillations of the $Q_{Ag}$ coordinate.[5,11,12]

As detailed in the main text, coherent non-polar $A_g$-symmetry modes can be observed by Raman scattering of a femtosecond probe pulse in the time delay dependent reflectivity changes. The coherent component of the time-resolved reflectivity $\Delta R(t)$, shown in Figure 1b of the main text, is shown for $YBa_2Cu_3O_{6.48}$ and $YBa_2Cu_3O_{6.65}$ together with their Fourier transforms in

Figures S1a-d. The frequencies of the observed modes at 3.7, 4.2 and 4.5 THz agree with continuous-wave Raman scattering experiments and theoretical predictions of the same compound.[11] The real space motions of these impulsively driven modes, which involve oscillation of the in-plane Cu atoms, are depicted in Figure S1e. The measured amplitude of these Raman-active modes scales quadratically with the amplitude of the optically driven phonons $Q_{drive}$, as shown in Figure S1f.

The results agree with the calculated response of the Raman-active modes, shown for $YBa_2Cu_3O_{6.5}$ in Figure S2, which are driven by the third-order nonlinear coupling to the resonantly driven apical oxygen phonons $\sim Q_{drive}^2 Q_{Ag}$.

Next, we consider the lattice dynamics induced by quartic-order terms in $V_{anharm}$. Here, the infrared-active $B_{1u}$ modes fulfill the symmetry requirements for bi-quadratic coupling $Q_{IR,i}^2 Q_{IR,j}^2$ and linear-cubic coupling $Q_{IR,i}^3 Q_{IR,j}$. Due to the selective resonant excitation of only one polar phonon mode, coupling between three or more different modes is neglected. The corresponding anharmonic term then becomes $V_{anharm} = f_1 Q_{drive}^2 Q_{IR,2}^2 + f_2 Q_{drive}^3 Q_{IR,2} + f_3 Q_{drive} Q_{IR,2}^3$. The bi-quadratic term leads to a parametric amplification of the coupled phonons, as becomes apparent from the equations of motion

$$\ddot{Q}_{drive} + \gamma \dot{Q}_{drive} + (\omega_{drive}^2 + 2f_1 Q_{IR,2}^2) Q_{drive} + 3f_2 Q_{drive}^2 Q_{IR,2} + f_3 Q_{IR,2}^3$$
$$= Z_{drive}^* Q_{drive} E_{field}$$

$$\ddot{Q}_{IR,2} + \gamma \dot{Q}_{IR,2} + (\omega_{IR,2}^2 + 2f_1 Q_{drive}^2) Q_{IR,2} + f_2 Q_{drive}^3 + 3f_3 Q_{drive} Q_{IR,2}^2 = 0.$$

This parametric amplification is characterized by an exponential scaling of the coupled mode $Q_{IR,2}$ as a function of the driven mode $Q_{drive}$ and a parametric resonance at $\omega_{drive} = 2 \cdot \omega_{IR,2}$. Fig. S3 reports comprehensive calculations involving coupling between all $B_{1u}$ modes. We scaled the obtained amplitudes by $Z_{B1u,i}^*$ to calculate the polarization induced by each individual mode and convolved the results with the 30-fs time resolution of the experiment. We find agreement with the experimental observations (see Figure S4a-d).

## S2. Doping dependence of the amplified Josephson Plasmon Polariton

The SHS measurements on the $YBa_2Cu_3O_{6.48}$ sample presented in the main text were complemented by measurements on two differently doped compounds, namely underdoped $YBa_2Cu_3O_{6.65}$ ($T_C$ = 67 K) and optimally doped $YBa_2Cu_3O_{6.92}$ ($T_C$ = 94 K). Figure S5 shows the coherent signal oscillations extracted from the raw data for all three doping levels, measured at 5 K temperature base temperature and 7 MV/cm peak electric field, together with their corresponding Fourier amplitude spectra. The set of nonlinearly coupled phonons (grey peaks) remains unchanged for all three doping levels, whereas the Josephson Plasma frequencies shift with increasing doping to higher frequencies, tracking the blue shift of the Josephson Plasma edges in the equilibrium superconducting states.[2,13] The corresponding frequencies are 2.5 THz and 2.8 THz in $YBa_2Cu_3O_{6.48}$ and $YBa_2Cu_3O_{6.65}$, respectively (see red peaks in Figure S5b,d). With respect to $YBa_2Cu_3O_{6.48}$, the amplitude of the coherent plasma oscillations decreased at higher-doped $YBa_2Cu_3O_{6.65}$ and disappeared at optimally doped $YBa_2Cu_3O_{6.92}$. This can be explained by the resonance condition $\omega_{drive} = \omega_{J1}(q_{plas}) + \omega_{J2}(-q_{plas})$ for the nonlinear coupling between the resonantly driven phonon and the Josephson plasmons at finite $q$, which can still be fulfilled in $YBa_2Cu_3O_{6.65}$ ($\omega_{J1}(0) = 2$ THz, $\omega_{J2}(0) = 15$ THz) but not in $YBa_2Cu_3O_{6.92}$ ($\omega_{J1}(0) = 7.5$ THz, $\omega_{J2}(0) = 30$ THz).[2,13]

## S.3 Temperature dependence of the amplified Josephson Plasmon Polariton and phonons

The time-resolved second harmonic intensity $\Delta I_{SH}(t)$ was measured in the $YBa_2Cu_3O_{6.48}$ and $YBa_2Cu_3O_{6.65}$ samples over a broad range of base temperatures. Figure S6a shows the oscillatory signal contributions in $YBa_2Cu_3O_{6.48}$ for three representative values of 5 K (red), 300 K (light red) and 440 K (grey), extracted by removing the EFISH contribution as described in the Materials and Methods Section. The corresponding Fourier spectra are shown in Figure

S6b. The amplitude of the 2.5-THz mode in $YBa_2Cu_3O_{6.48}$ was found to extend far above the equilibrium critical temperature $T_c$ (~ 45 K), and to vanish only above T = 400 K. The same measurements allowed to also quantify the temperature dependent amplitude of the amplified phonons $Q_{amplified}$, by integrating the area under the Fourier transformations in the spectral range between 7 and 11 THz.

The full temperature dependences of the amplified Josephson Plasmon Polaritons and the amplified phonons, shown in Figure 2h,i of the main text, were obtained by repeating the temperature dependent SHS measurements several times and calculating the mean values of the amplitudes to account for systematic errors due to sample drift.

## S4. SHS polarimetry

To describe the SHS polarimetry signals of the symmetry-odd modes in Figure 3, we consider that the instantaneous second harmonic intensity, which is mediated by a third-order susceptibility $\chi^3(\omega_{IR}, \omega_{IR}, \omega_{THz})$, can be described as an effective second-order optical nonlinearity $\chi^2_{eff}(\omega_{IR}, \omega_{IR}, t)$. Hence, the polarization angle dependence can be calculated by considering the full tensorial form of this effective second harmonic generation process

$$P_i^{2\omega} = \sum_{j,k} \chi^{(2)}_{ijk} E_j^\omega E_k^\omega . \qquad (4.1)$$

Here, the indices $i,j,k$ represent the polarization directions of the polarization $P$ and the electric fields $E$, where the directions 1,2,3 correspond to the crystal axes $a,b,c$ of the $YBa_2Cu_3O_{6+x}$ unit cell[43].

The nonlinear coefficient $\chi^{(2)}_{ijk}$ is a third-rank tensor that connects the three interacting fields. Given the Kleinman symmetry condition, the number of independent elements of this tensor can be reduced to 18 elements $d_{il}$ and the most general equation describing second harmonic generation is

$$\begin{pmatrix} P_1^{2\omega} \\ P_2^{2\omega} \\ P_3^{2\omega} \end{pmatrix} = 2 \begin{pmatrix} d_{11} & d_{12} & d_{13} & d_{14} & d_{15} & d_{16} \\ d_{21} & d_{22} & d_{23} & d_{24} & d_{25} & d_{26} \\ d_{31} & d_{32} & d_{33} & d_{34} & d_{35} & d_{36} \end{pmatrix} \begin{pmatrix} E_1^2 \\ E_2^2 \\ E_3^2 \\ 2E_2 E_3 \\ 2E_1 E_3 \\ 2E_1 E_2 \end{pmatrix}. \quad (4.2)$$

Depending on the point group symmetry, the number of independent tensor elements can be further reduced. Given the symmetries of the amplified optical phonons and the Josephson Plasmon Polaritons (see Fig. 5f,g of the main text), the relevant point groups for the discussion in this work are *mm2* and *m*. Their corresponding $d_{il}$ tensors read

$$d_{il}(mm2) = \begin{pmatrix} 0 & 0 & 0 & 0 & d_{15} & 0 \\ 0 & 0 & 0 & d_{24} & 0 & 0 \\ d_{31} & d_{32} & d_{33} & 0 & 0 & 0 \end{pmatrix} \quad (4.3)$$

and

$$d_{il}(m) = \begin{pmatrix} d_{11} & d_{12} & d_{13} & 0 & d_{15} & 0 \\ 0 & 0 & 0 & d_{24} & 0 & d_{26} \\ d_{31} & d_{32} & d_{33} & 0 & d_{35} & 0 \end{pmatrix}. \quad (4.4)$$

By choosing an appropriate orientation of the analyzer in front of the detector two orthogonal polarization components $P_i^{2\omega}$ can be measured. In our experimental geometry, the two accessible components were $P_3^{2\omega}$ (s-analyzer) and $P_1^{2\omega} + P_2^{2\omega}$ (p-analyzer). Equation 4.2 can then be used to fit the shape of the SHS polarimetry signals and determine the individual tensor elements $d_{il}$ at any given time delay.

The results of the SHS polarimetry measurements with the analyzer oriented along the YBa$_2$Cu$_3$O$_{6.48}$ *c* axis (s-analyzer) are shown in Figure 3 of the main text. The measurements taken with the analyzer oriented along the *b* axis (p-analyzer, see the optical setup in Figure S7a) are discussed here.

Figure S7b shows the corresponding measurement as a function of incoming 800-nm polarization angle $\varphi$ and pump probe time delay. The frequency-filtered SHS polarimetry signals for one representative time delay ($t = 500$ fs) are shown in Figures S7c,d,e. Again, the directly driven phonons (yellow) and the amplified phonons (grey) can be fitted by the $\chi^2$ tensor of the *mm2* point group (dashed lines), as expected for the $B_{1u}$-symmetry lattice distortions. Also, the sign of the phonon amplitudes at this delay, and hence their phases, are independent of the polarization angle $\varphi$. In contrast, the $\varphi$ angular dependence of the 2.5 THz Josephson plasmon mode, shown in Figure S7e, requires a fit by the $\chi^2$ tensor of a lower-symmetry point group (*m* or lower). In addition, the amplitude of this mode changes sign as a function of incoming polarization angle $\varphi$.

Both results agree with the symmetry analysis presented in the main text for the s-analyzer configuration.

## S5. Momentum-resolved detection of the Josephson Plasmon Polariton

In the SHS measurement, the 400-nm wavelength light is generated in a thin layer $l$ of about 100 nm below the sample surface. The finite in-plane momentum $q_y$ of the amplified Josephson Plasmon Polariton leads to a deflection of the second harmonic light with respect to the specular reflection. The spatial distribution of the emitted radiation was determined by taking second harmonic intensity measurements $\Delta I_{SH}(t)$ at different positions of a 200-µm slit, which was scanned across the re-collimated reflected beam. The amplitudes of the frequency-filtered 2.5-THz JPP and the amplified phonon contributions are plotted as a function of the slit position in Figure S8b,c, together with the EFISH amplitude at time zero shown in Figure S8a. The momentum transfer was calculated from the deflection angle $\Delta\theta$ by

$$q_y = \sqrt{\varepsilon_\infty}\, tan(\Delta\theta) k_{400},$$

where $k_{400}$ is the vacuum wavenumber of the 400-nm light. While both the EFISH and amplified phonon contributions are symmetric and peak at zero in-plane momentum transfer $q_y$, the plasmon response is asymmetric and peaks at a finite momentum $q_y$ = 190 cm$^{-1}$.

The momentum distribution of the Josephson Plasmon Polariton, shown in Fig. 4d of the main text, was then obtained by deconvolving the measured JPP profile (Figure S8c) from the divergence of the probe beam. To this end, consistent with our theoretical model, two constrained Gaussian profiles, with equal but opposite abscissa offsets and same widths, were fitted to the data. The best fit was deconvolved with the Gaussian profile of the undeflected second harmonic beam, which is accessible from the momentum dependent EFISH signal due to the collinearity between the incident mid-IR excitation and 800-nm probe beams in these measurements. The error-bars of the deconvoluted data points are determined by the deviation of the least squares fit to the data points. Note that while the deconvolution result is not unique, the 190-cm$^{-1}$ momentum shift is already clearly visible in the raw data (Figure S8c).

**S6. Excitation frequency dependence of the Josephson Plasmon Polariton amplification**

Our theory predicts the amplification of the Josephson Plasmon Polariton by three-wave mixing with the *c*-axis apical oxygen phonon mode. Hence, we expect the amplification to be enhanced when the mid-infrared excitation pulses are frequency-tuned into resonance with this phonon, where the latter is driven to largest amplitudes.

We tested this prediction by recording the amplitude of the 2.5 THz mode in YBa$_2$Cu$_3$O$_{6.48}$ for different center frequencies of the mid-infrared pulses, keeping the peak electric field constant at ~7 MV/cm. In Figure S9, we plot this dependence together with the real part of the optical conductivity. Clearly, the Josephson Plasmon Polariton amplitude increases when the mid-infrared pulses are tuned into the phonon resonance, supporting the proposed three-wave phonon-plasmon mixing.

Furthermore, Figure S9 shows that the frequency of the amplified Josephson Plasmon Polariton does not change as function of the mid-infrared center frequency. Given the JPP dispersion, this implies that in the three-wave mixing process, with resonance condition $\omega_{IR} = \omega_{J1}(q_{plas}) + \omega_{J2}(-q_{plas})$, the frequency $\omega_{IR}$ always takes the same value. Hence, this has to be the eigenfrequency $\omega_{drive}$ of the phonon and not the tunable frequency of the excitation pulses.

Together, these two observations show that the amplified JPP amplitude scales with the resonant enhancement of the driven apical oxygen phonon amplitude $Q_{drive}$ and exclude a scenario, where the incident light field couples directly to the Josephson Plasmon Polariton.

**S7. Theoretical analysis of the Josephson Plasma Polariton**

*Analysis of the collective modes*

Plasmon dispersion in a layered superconductor can be obtained by combining linearized dynamical equations for superflow currents and charges with Maxwell equations for electromagnetic fields.[24,25,44,45,46,47] The fundamental degrees of freedom are density fluctuations of the condensate $\delta\rho_{\lambda,i}(\vec{x})$, the phase of the superconducting order parameter $\phi_{\lambda,i}(\vec{x})$, and the 4-component vector potential $(V_{\lambda,i}(\vec{x}), A_{\lambda,i,z}(\vec{x}) \vec{A}_{\lambda,i,\vec{x}}(\vec{x}))$. Here $i$ corresponds to the index of the unit cell along the c-axis, $\lambda = 1,2$ labels the number of the layer inside the unit cell, and $\vec{x}$ is the in-plane coordinate, which we will omit in the equations below for brevity. While the in-plane components of the vector potential $\vec{A}_{\lambda,i,\vec{x}}(\vec{x})$ are defined within the corresponding layers, $A_{\lambda,i,z}(\vec{x})$ is defined to be on the links between layers starting on layer $\{\lambda, i\}$ as shown in Fig. S10.

In linearized hydrodynamics, superflow currents are given by

$$j_{\lambda,i,\vec{x}} = \Lambda_s(\partial_{\vec{x}}\phi_{\lambda,i} - e^*A_{\lambda,i,\vec{x}}), \quad (6.1)$$

$$j_{\lambda,i,z} = j_{c\lambda}(\Delta_z\phi_{\lambda,i} - e^*A_{\lambda,i,z}). \quad (6.2)$$

Here $\vec{x}$ denotes the in-plane $x, y$ components and $z$ denotes the c-axis coordinate of the crystal. Coupling to the vector potential is given by the Cooper pair charge, $e^* = 2e$, and we work in units where $\hbar = 1$ for the rest of this section. The in-plane components of the superfluid current are defined within individual layers and have continuous gradients. The $z$ component of the current is defined as the Josephson current between adjacent layers and has a lattice gradient which corresponds to the phase difference between adjacent layers,

$$\Delta_z \phi_{\lambda,i} = \begin{cases} (\phi_{2,i} - \phi_{1,i})/d_1, \text{ for } \lambda = 1, \\ (\phi_{1,i+1} - \phi_{2,i})/d_2, \text{ for } \lambda = 2 \end{cases} \quad (6.3)$$

The coefficient $\Lambda_s$ is related to the in-plane London penetration length as $\Lambda_s = \frac{\epsilon c^2}{\lambda_L^2 (e^*)^2}$, where $\epsilon = \epsilon_r \epsilon_0$. Physically, it corresponds to the intra-layer superfluid stiffness and is proportional to the condensate density, $\Lambda_{s\lambda} \propto \rho_\lambda$. In linear analysis of the collective modes we can set $\Lambda_{s\lambda}$ to be equal to their equilibrium values since they multiply superfluid velocities, $\vec{v}_{\lambda,i} = \partial_{\vec{x}} \phi_{\lambda,i} - e^* A_{\lambda,i,\vec{x}}$, which are already first order in fluctuations. This is why we omitted the layer index for $\Lambda_s$ in equation (1). Coefficients $\{j_{c,\lambda}\}$ correspond to interlayer Josephson tunneling couplings and obey $j_{c,\lambda} \propto \sqrt{\rho_1 \rho_2}$. In linearized hydrodynamics we take $j_{c,\lambda}$ to be equal to their equilibrium value and neglect corrections due to $\delta \rho_\lambda$. Both $\rho_\lambda$ and $j_{c,\lambda}$ can be modified by exciting apical oxygen phonons, which results in phonon-plasmon coupling that will be discussed below. We introduce an effective Hamiltonian that describes plasmon degrees of freedom and show that its equations of motion give the correct equations for light and matter fields.

$$H = H_{pot.} + H_{kin.} + H_{EM.} \quad (6.4)$$

The first term in eqn. (4) describes finite compressibility of Cooper pairs and their coupling to electrostatic potential

$$H_{pot.} = \int d^2x \sum_{i,\lambda} \left\{ \frac{\gamma}{2} \delta \rho_{\lambda,i}^2 + e^* \delta \rho_{\lambda,i} V_{\lambda,i} \right\}. \quad (6.5)$$

Compressibility $\gamma$ can be related to the Thomas-Fermi length, $\lambda_{TF}$, as $\gamma = \frac{\lambda_{TF}^2 (e^*)^2}{\epsilon}$.

The superflow kinetic energy is given by

$$H_{kin.} = \int d^2x \sum_{i,\lambda} \left\{ \frac{1}{2\Lambda_s} j_{\lambda,i,\vec{x}}^2 + \frac{1}{2j_{c,\lambda}} j_{\lambda,i,z}^2 \right\}. \tag{6.6}$$

For electromagnetic fields we adopt the Lorenz gauge condition

$$\frac{1}{c^2}\partial_t V_{\lambda,i} + \partial_{\vec{x}} A_{\lambda,i,\vec{x}} + \Delta_z A_{\lambda,i,z} = 0 \tag{6.7}$$

then the Hamiltonian for electromagnetic fields is given by

$$\begin{aligned}
H_{EM} = \int d^2x \Bigg\{ & \sum_{i,\lambda} \frac{c^2}{2\epsilon} P_{V,\lambda,i}^2 + \frac{\epsilon}{2}\left( (\partial_{\vec{x}} V_{\lambda,i})^2 + (\Delta_z V_{\lambda,i})^2 \right) \\
& + \frac{1}{2\epsilon} P_{A_{\vec{x}},\lambda,i}^2 + \frac{\epsilon c^2}{2}\left( (\partial_{\vec{x}} A_{\lambda,i,\vec{x}})^2 + (\Delta_z A_{\lambda,i,\vec{x}})^2 \right) \\
& + \frac{1}{2\epsilon} P_{A_z,\lambda,i}^2 + \frac{\epsilon c^2}{2}\left( (\partial_{\vec{x}} A_{\lambda,i,z})^2 + (\Delta_z A_{\lambda,i,z})^2 \right) \Bigg\}
\end{aligned} \tag{6.8}$$

Variables $\{P_{V,\lambda,i}, P_{A_{\vec{x}},\lambda,i}, P_{A_z,\lambda,i}\}$ correspond to the conjugate momenta of the scalar and vector potentials, and magnetic permeability $\mu = \mu_r \mu_0$ is included in the speed of light $c^2 = 1/\mu\epsilon$. In eqn. (8) gradients in the $z$ direction are taken in the lattice form so, for example,

$$\Delta_z A_{\lambda,i,z} = \begin{cases} \dfrac{A_{1,i,z}}{d_1} - \dfrac{A_{2,i-1,z}}{d_2}, & \text{for } \lambda = 1, \\ \dfrac{A_{2,i,z}}{d_2} - \dfrac{A_{1,i,z}}{d_1}, & \text{for } \lambda = 2 \end{cases} \tag{6.9}$$

We use Heisenberg equations of motion (EOM) for the operators, $\partial_t \hat{O} = i[H, \hat{O}]$, to study dynamics of the fields. In deriving equations of motion we use canonical commutation relations between $\rho$ and $\phi$, $V$ and $P_V$, $\vec{A}$ and $P_{\vec{A}}$, i.e. $[\rho_i(\vec{x}), \phi_j(\vec{x}')] = i\delta^2(\vec{x} - \vec{x}')\delta_{i,j}$, etc. EOM for the density and phase operators give the continuity equation and Josephson relation

$$\partial_t \delta\rho_{\lambda,i} + \partial_{\vec{x}} j_{\lambda,i,\vec{x}} + \Delta_z j_{\lambda,i,z} = 0. \tag{6.10}$$

$$\partial_t \phi_{\lambda,i} = -\gamma \delta\rho_{\lambda,i} - e^* V_{\lambda,i}, \qquad (6.11)$$

By combining EOM for the electromagnetic fields $\phi, \vec{A}$ and their conjugate momenta we obtain Maxwell's equations:

$$\left(\frac{1}{c^2}\partial_t^2 - \partial_{\vec{x}}^2 - \Delta_z^2\right) V_{\lambda,i} = \frac{e^*}{\epsilon} \delta\rho_{\lambda,i}, \qquad (6.12a)$$

$$\left(\frac{1}{c^2}\partial_t^2 - \partial_{\vec{x}}^2 - \Delta_z^2\right) A_{\lambda,i,\vec{x}} = \frac{1}{c^2}\frac{e^*}{\epsilon} j_{\lambda,i,\vec{x}}, \qquad (6.12b)$$

$$\left(\frac{1}{c^2}\partial_t^2 - \partial_{\vec{x}}^2 - \Delta_z^2\right) A_{\lambda,i,z} = \frac{1}{c^2}\frac{e^*}{\epsilon} j_{\lambda,i,z} \qquad (6.12c)$$

To find collective modes we look for the solutions of equations (10), (11), (12) in the form of plane waves, $\delta\rho_{\lambda,l}(x,t) = \delta\rho_\lambda(q_{\vec{x}}, q_z, \omega) e^{i(q_x x + q_y y + q_z D l - \omega t)}$, with similar expressions for other variables. It is convenient not to combine EOM for electromagnetic fields and their conjugate variables, so that we have first order linear differential equations of the form $\partial_t \vec{v} = \underline{M} \vec{v}$. Matrix $\underline{M}$ contains gradient operators which leads to implicit dependence on momentum $\vec{q}$. We define the characteristic polynomial for $\underline{M}$ as $\chi(\omega) = \det |i\omega + \underline{M}|$. Due to the Lorenz gauge used in our analysis the characteristic polynomial contains unphysical degrees of freedom. However, gauge constraint (7) guarantees that they do not couple to matter fields and the characteristic polynomial factorizes into physical and unphysical contributions, $\chi(\omega) = \chi_{phys}(\omega)\chi_{unphys}(\omega)$. Collective modes of the system can be found by solving the secular equation $\chi_{phys}(\omega) = 0$. The two lowest energy modes correspond to the Josephson plasmons and their dispersion is shown in Fig. S11.

To express physical quantities in terms of the amplitudes of the plasmon modes we can use eigenvectors $v^l_{\{1,2\},q}$ of the secular equation, where components $l$ correspond to $\delta\rho_\lambda, \phi_\lambda, V, \vec{A}$, etc. Shown in a matrix form:

$$\begin{pmatrix} \rho_\lambda(q) \\ \vdots \\ \phi_\lambda(q) \\ \vdots \end{pmatrix} = \begin{pmatrix} v_{1,q}^{\delta\rho_\lambda} & \cdots & (v_{1,q}^{\delta\rho_\lambda})^* & \cdots \\ \vdots & & \vdots & \\ v_{1,q}^{\phi_\lambda} & & (v_{1,q}^{\phi_\lambda})^* & \\ \vdots & & \vdots & \end{pmatrix} \cdot \begin{pmatrix} b_1 \\ \vdots \\ b_1^* \\ \vdots \end{pmatrix} \tag{6.13}$$

where $b_1$ and $b_2$ are amplitudes of the two plasmon modes oscillating at frequencies corresponding to their dispersion relations. The eigenvectors, $v_{\{1,2\},q}^l$, are defined through the EOM up to a normalization constant. Normalization is fixed through the commutation relations of canonically conjugate pairs, such as $[\rho_\lambda(q), \phi_{\lambda'}(q')] = i\delta_{q,q'}\delta_{\lambda,\lambda'}$, and commutation relations of the plasmon fields, which should correspond to bosonic creation/annihilation operators $[b_i, b^\dagger] = \delta_{i,j}$.

*Phonon-plasmon interaction*

The apical oxygen phonon is expected to modify the in-plane superfluid stiffness either by changing the in-plane density of carriers or by modifying their hopping. Symmetry of this mode requires that these changes are antisymmetric with respect to the two layers inside one unit cell, so that $\delta\Lambda_{s,\{1,2\}}^{phon}(t) = \pm\xi Q_{drive}(t)\Lambda_s$, where coefficient $\xi$ characterizes the coupling strength. Changes of the interlayer Josephson currents arise from changes in the superfluid density $\delta\rho_{\{1,2\}}^{phon} = \pm\tilde{\xi} Q_{drive}(t)\rho$, which results in $\delta j_{c,\lambda}(t) = -\left(\tilde{\xi} Q_{IR}\right)^2 \frac{j_{c,\lambda}}{2\rho}$. The last equation shows that interlayer Josephson currents couple quadratically to the apical oxygen phonon and lead to four-wave phonon/plasmon mixing. Resonant three wave mixing considered in the main text comes from phonons modifying $\Lambda_{s,\{1,2\}}$ and coupling to the in-plane current.

To derive plasmon dynamics in the presence of excited phonon mode we need to modify equation (6) to include $\delta\Lambda_{s\lambda}^{phon}$ arising due to phonons. We find

$$\delta H_{kin.} = -\xi \sum_i \int d^2x \left\{ \frac{Q_{IR}(t)}{2\Lambda_s} \left( j^2_{1,i,\vec{x}} - j^2_{2,i,\vec{x}} \right) \right\} \tag{6.14}$$

The phonon mode causes a zero momentum three wave parametric process that excites pairs of plasmons at opposite momenta. Resonant processes that satisfy energy matching condition $\omega_{drive} = \omega_1(q) + \omega_2(-q)$ lead to exponential instability discussed in the main text. After projecting the modified EOM to the two the lowest eigenmodes we find equations for parametrically coupled oscillators

$$\partial_t^2 J_1(q) + 2\gamma_1 \partial_t J_1(q) + \omega_1^2(\vec{q}) J_1(q) = -q_x^2 f(\vec{q}) Q_{drive}(t) J_2(q), \tag{6.15a}$$

$$\partial_t^2 J_2(q) + 2\gamma_2 \partial_t J_2(q) + \omega_2^2(\vec{q}) J_2(q) = -q_x^2 f(\vec{q}) Q_{drive}(t) J_1(q) \tag{6.15b}$$

In writing equations (15) we added phenomenological damping constants $\gamma_i$ to describe dissipation due to quasiparticles. Factors of $q_x^2$ in equations (15) originate from the fact that phonons couple to plasmons through the in-plane superflow kinetic energy. There is also an implicit weaker $q$ dependence in $f(\vec{q})$ arising from projecting the interaction to the plasmon subspace, which can be derived using the $v^{j\lambda\vec{x}}_{\{1,2\},q}$ components of the eigenvectors of the secular equation. We also note that inversion symmetry forbids three-mode coupling between the phonon and Josephson plasmons in the same band.

The equation of motion for the polar phonon $Q_{drive}$ reads

$$\ddot{Q}_{drive} + 2\gamma_{drive} \dot{Q}_{drive} + \omega_{drive}^2 Q_{drive} = Z^* E(t) - q_y^2 J_1 J_2,$$

where $Z^*$ is the coupling to the optical drive field and $\gamma_{drive}$ accounts for the finite lifetime of the vibrational mode.

We solved the set of coupled equations for the phonon and plasmon dynamics by utilizing a stochastic approach, where we introduced Langevin noise on both, the Josephson plasma and phonon coordinates, to create an incoherent initial state. The final trajectories were then computed by solving the equations of motion one million times with an algorithm based on the Euler-Maruyama method. In addition to the harmonic terms that describe the resonant driving of the polar phonon mode, we also considered higher-order (quartic) phonon anharmonicities.

## Materials and Methods

### Sample preparation

The samples are detwinned single crystals of $YBa_2Cu_3O_{6+\delta}$ grown in Y-stabilized zirconium crucibles. The hole doping of the Cu-O planes was adjusted by controlling the oxygen content of the CuO chain layer through annealing in flowing $O_2$ and subsequent rapid quenching.[48] The critical temperatures of the superconducting transitions were determined by dc magnetization measurements ($T_c$ = 45 K for $YBa_2Cu_3O_{6.48}$, $T_c$ = 67 K for $YBa_2Cu_3O_{6.65}$ and $T_c$ = 94 K for $YBa_2Cu_3O_{6.92}$).

For the optical experiments, *bc*-surfaces of the single crystals were polished to optical grade with a final lapping step at 100 nm grid size. The samples were then mounted into an optical cryostat where their temperature could be controlled between 10 and 450 K.

### Optical setup

The carrier envelope phase (CEP) stable mid-infrared pump pulses were obtained by mixing the two signal beams from two optical parametric amplifiers, which were seeded by the same white light and pumped by 30-fs pulses at 800 nm wavelength and 1 kHz repetition rate. The pulses were 150 fs long and centered at 17.5 THz with a bandwidth of 5 THz. The driven dynamics in the $YBa_2Cu_3O_{6+x}$ samples were probed by time-delayed replicas of the 800-nm pulses, in non-collinear geometry with an angle of ~17° to the normal-incidence mid-infrared pump (see Fig. S12). Only for the measurements of the nonlinear scattering angle in the second harmonic generation, the two beams were aligned collinearly onto the sample.

The pump induced polarization rotation of the 800-nm pulses, reflected from the $YBa_2Cu_3O_{6+x}$ samples, was measured by detecting the time-resolved difference signal of two intensity-balanced photodiodes placed behind a half-wave plate and a Wollaston prism.

The light fields generated at the second harmonic frequency (SH) at 400 nm wavelength were separated from the fundamental beam behind the sample by a dichroic mirror and then detected by a photo multiplier tube. For SH polarimetry measurements, the polarization angle of the incoming 800-nm wavelength pulses was rotated by a half-wave plate. In this case, a polarizer was placed as an analyzer in front of the photomultiplier tube to detect the 400-nm SH intensity with polarization either along the c axis or the b axis of the $YBa_2Cu_3O_{6.48}$ single crystal. Both schemes probe the material up to a depth of approximately ~0.1 μm, much smaller than the penetration depth of the mid-infrared excitation of about 1.5 μm.[1,2,3]

**Analysis of the time-resolved SHS signals**

The measured time-resolved second harmonic intensity $\Delta I_{SH}(t)$ (see for example Fig. 1c of the manuscript) was fitted by the product of (i) a Gaussian envelope to map the nonlinear optical mixing of pump and probe electric fields at time delay zero (electric field induced second harmonic generation, EFISH) and (ii) a step function of finite width, multiplied by a decaying exponential, i.e. $A(\tau) = A \cdot (1 + \text{erf}((\tau - \tau_0)/\sigma)) \cdot \exp(-\gamma(\tau - \tau_0))$, to describe the exponentially decaying background signal. Subtraction of this slowly varying background revealed the coherent oscillations shown in Figures 2a, 2c and e of the main text for different excitation strengths. The oscillatory signals can be divided into sets of three exponentially decaying oscillators: the driven polar phonons $Q_{drive}$, the amplified phonons $Q_{amplified}$, and the nonlinearly coupled Josephson Plasmon Polaritons $J_1$ and $J_2$ (see Fig. S4). These were fitted as $A(\tau) = A \cdot (1 + \text{erf}((\tau - \tau_0)/\sigma)) \cdot \exp(-\gamma(\tau - \tau_0)) \cdot \sin(2\pi\Omega\tau + \phi_0)$. For the phonons, their frequencies $\Omega_{phonon}$ were constrained to values measured by linear infrared spectroscopy.[19]

# Supplementary Figures

**Experiment: Raman Phonons**

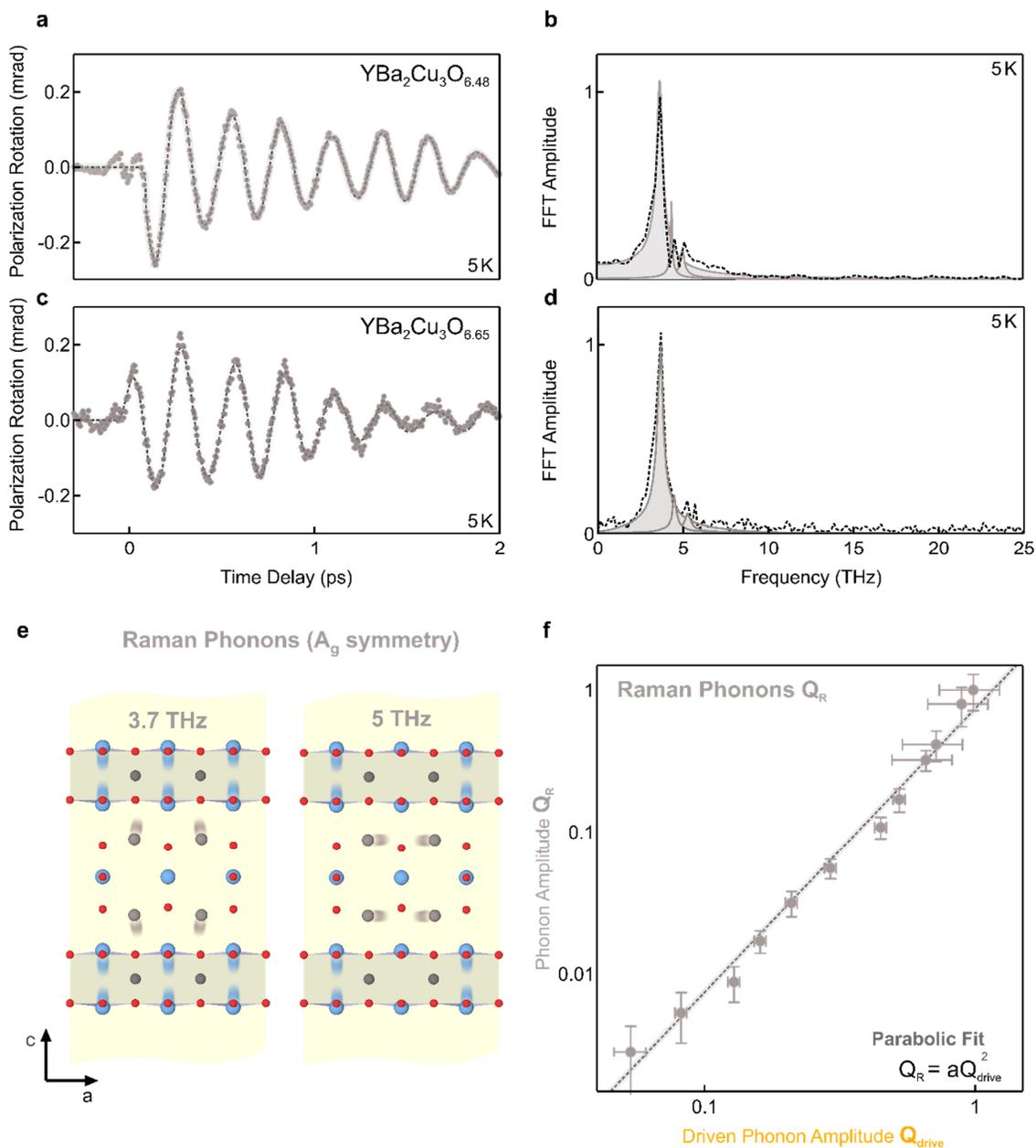

**Fig. S1:** Coherent component of the polarization rotation signal of $YBa_2Cu_3O_{6.48}$ in panel **a** and of $YBa_2Cu_3O_{6.65}$ in panel **c**. The dashed lines are a time domain fit with several exponentially decaying oscillators. The corresponding Fourier amplitude spectra are shown in panel **b** and **d**, respectively. The individual oscillators of the time domain fits are shown as grey peaks. **e** Real space motion of the two dominant Raman modes at 3.7 and 5 THz determined by ab-initio

methods. **f**, Measured excitation strength dependence of the amplitude of the 3.7 THz mode as a function of the amplitude of the driven apical oxygen vibration The dashed black line is a parabolic fit, $Q_R = aQ_{drive}^2$. Error bars represent the standard deviation $\sigma$ of the amplitudes extracted by numerical fits.

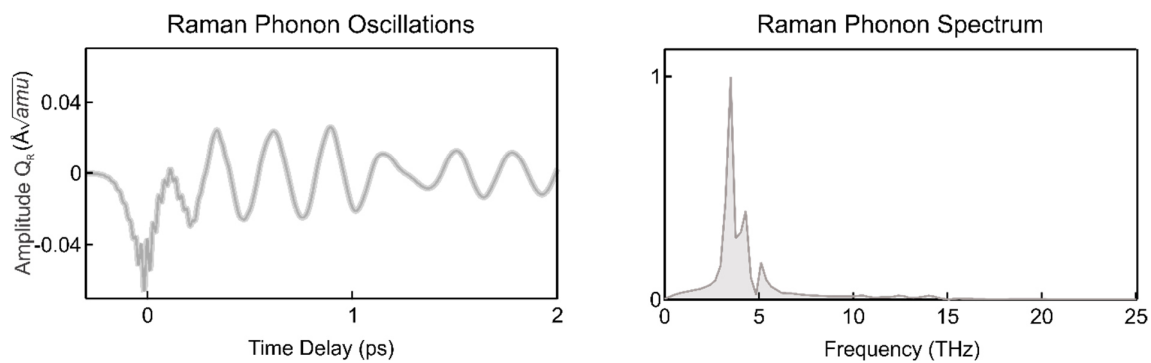

**Fig. S2:** Simulated coherent oscillations of the Raman modes excited by ionic Raman scattering due to their third order nonlinear coupling to the resonantly driven apical oxygen phonons. The corresponding frequency spectrum is shown in the right panel and displays the same dominant response at 3.7 and 5 THz as observed in the experiment.

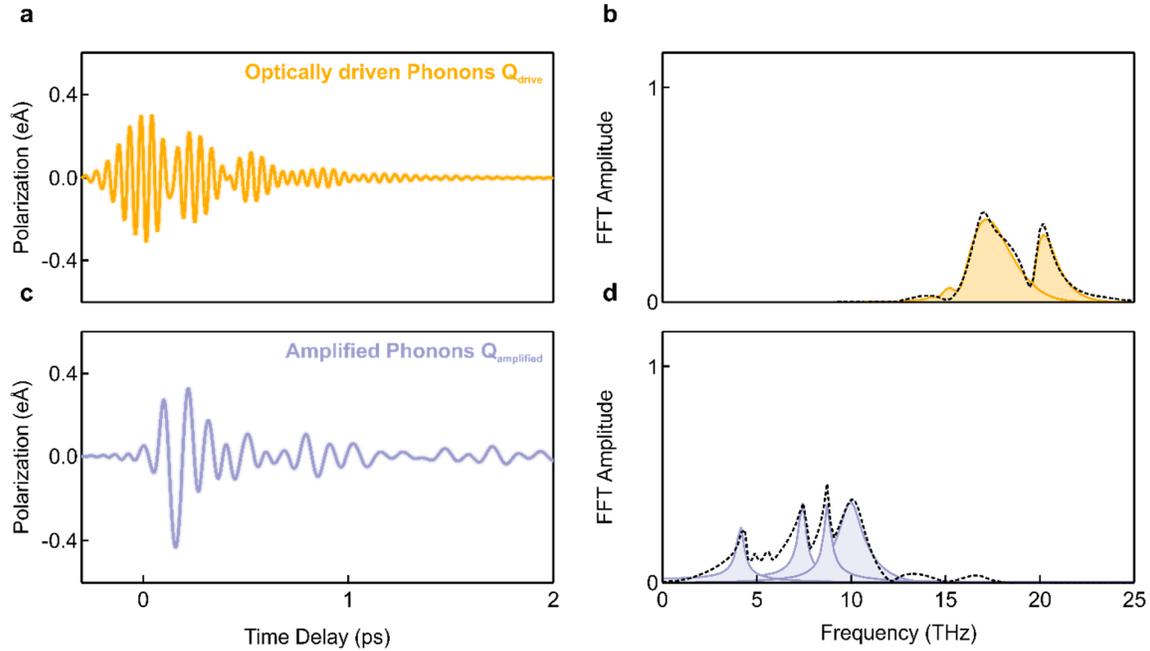

**Fig. S3: a** Simulated coherent oscillations of the resonantly driven apical oxygen vibrations $Q_{drive}$ (yellow). **b**, The corresponding Fourier amplitude spectrum exhibits two peaks at 17 and 20 THz (dashed line). The individual phonon responses are shown as yellow peaks, whilst the total response is drawn as a dashed black line. **c**, Simulated coherent oscillations of the nonlinearly coupled (amplified) infrared-active vibrations $Q_{amplified}$ at 4, 6, 8 and 10 THz (grey), which are driven through fourth-order nonlinear coupling to the resonantly driven apical oxygen phonons. **d**, The Fourier amplitude spectrum (dashed black) can be dissected into the response of the individual modes, which are shown as shaded grey peaks. The response is similar to the experiment.

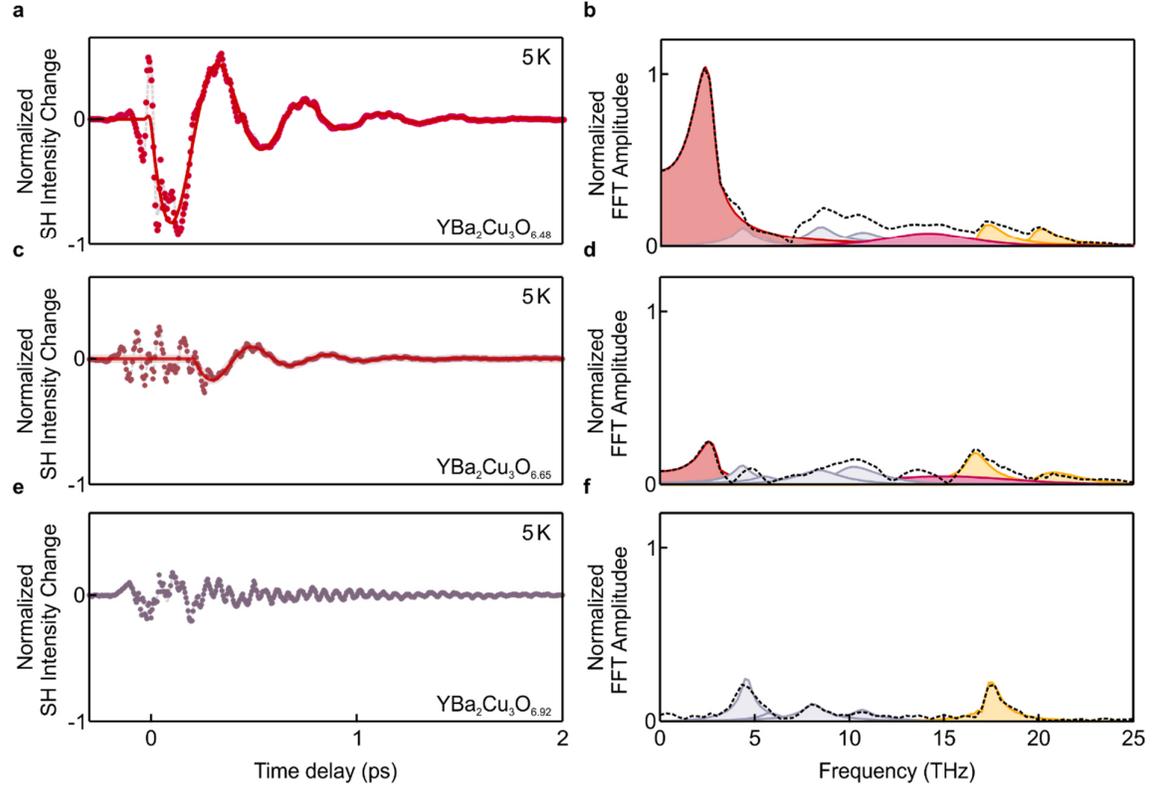

**Fig. S4:** Coherent oscillations in the time-resolved second harmonic intensity $\Delta I_{SH}(t)$ from YBa$_2$Cu$_3$O$_{6.48}$, as shown in Fig. 2e of the main text for 7MV/cm excitation at 5 K, divided into three contributions and shown together with their Fourier amplitude spectra. Panels **a,b** show the driven apical oxygen phonons, panels **c,d** show the amplified infrared-active phonons, and panels **e,f** show the Josephson Plasma Polariton modes. Experimental phonon oscillations (yellow and grey dots in a and c, respectively) are fitted by oscillators with frequencies constrained by infrared spectroscopy data (dashed lines).[19] Oscillations of the two Josephson plasma modes (experimental data as red dots in e) are best fitted by two oscillators at 2.5 and 14 THz (dashed line). In the Fourier amplitude spectra, the colored peaks highlight the respective contributions.

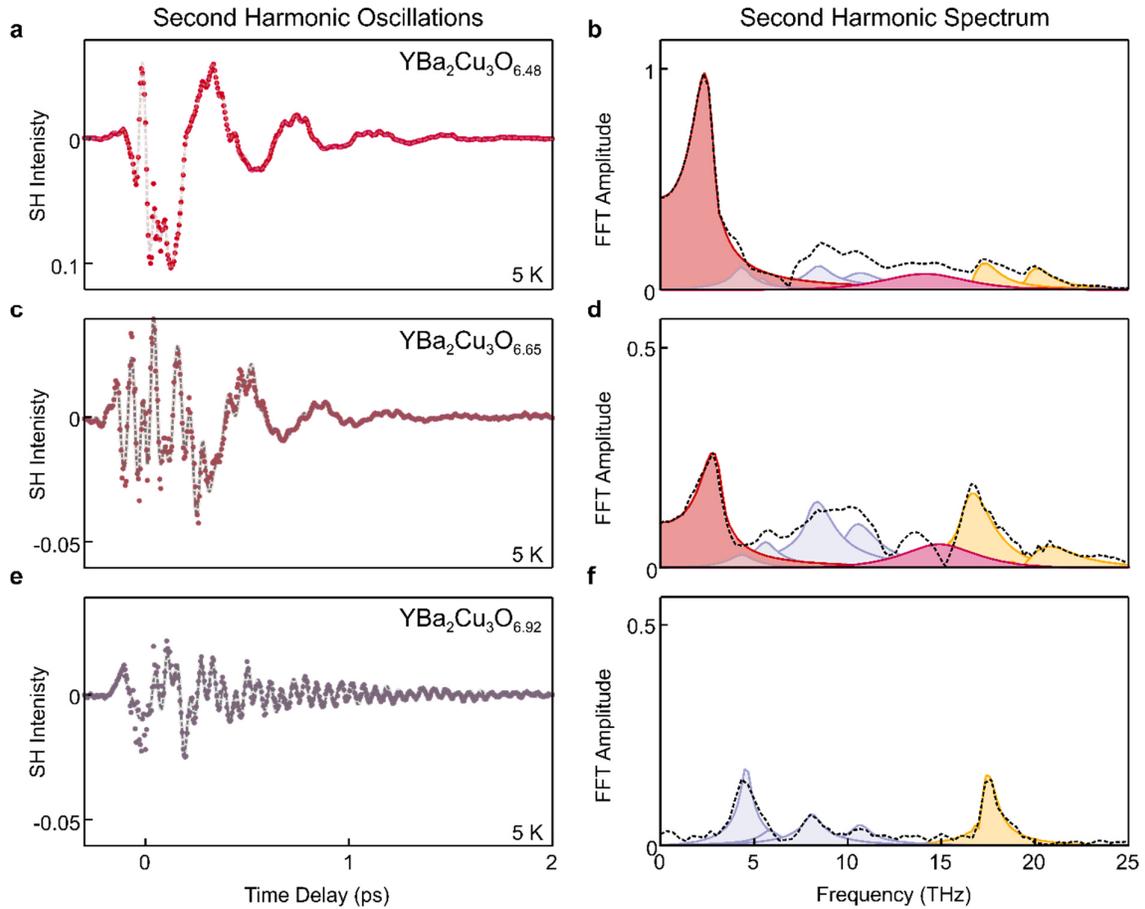

**Fig. S5** Coherent oscillations in the time delay dependent second harmonic intensity of **a** YBa$_2$Cu$_3$O$_{6.48}$, **c** YBa$_2$Cu$_3$O$_{6.65}$ and **e** YBa$_2$Cu$_3$O$_{6.92}$, measured in the superconducting state at 5 K, together with corresponding Fourier amplitude spectra in **b, d, f**. Experimental data are plotted as red dots in panels a, c, e, together with the best fits to the data (light grey dashed lines) and the dominating low-frequency Josephson plasmon contribution (black solid line). The Fourier amplitude spectra include Josephson plasmons as red and magenta peaks, resonantly driven apical oxygen phonons as yellow peaks, and the nonlinear coupled infrared active phonons as grey peaks. Note the absence of the Josephson plasma oscillations in YBa$_2$Cu$_3$O$_{6.92}$. All experiments were performed with a peak electric field strength of 7 MV/cm.

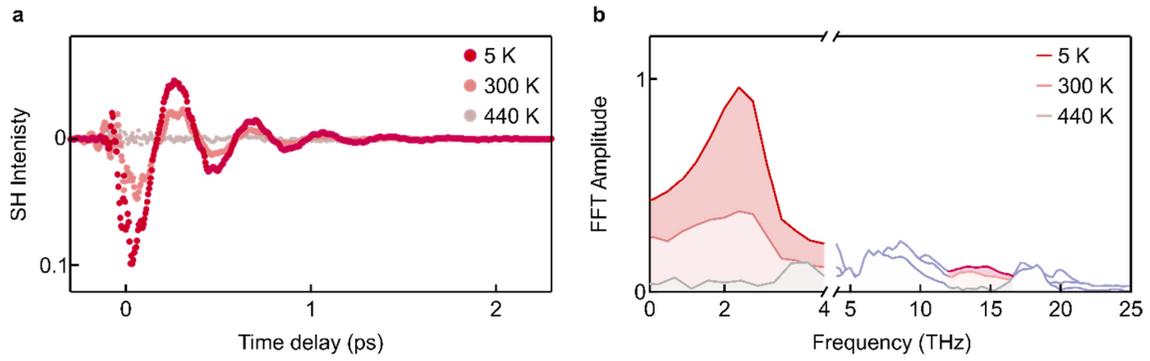

**Fig. S6 a** Temperature dependent coherent oscillations of the SH intensity in $YBa_2Cu_3O_{6.48}$ for three temperatures base temperatures 5 K (red), 300 K (light red) and 440K (grey). Their corresponding FFT amplitude spectra are shown in panel **b**. The shaded areas highlight the spectral regions of the amplified Josephson plasma modes.

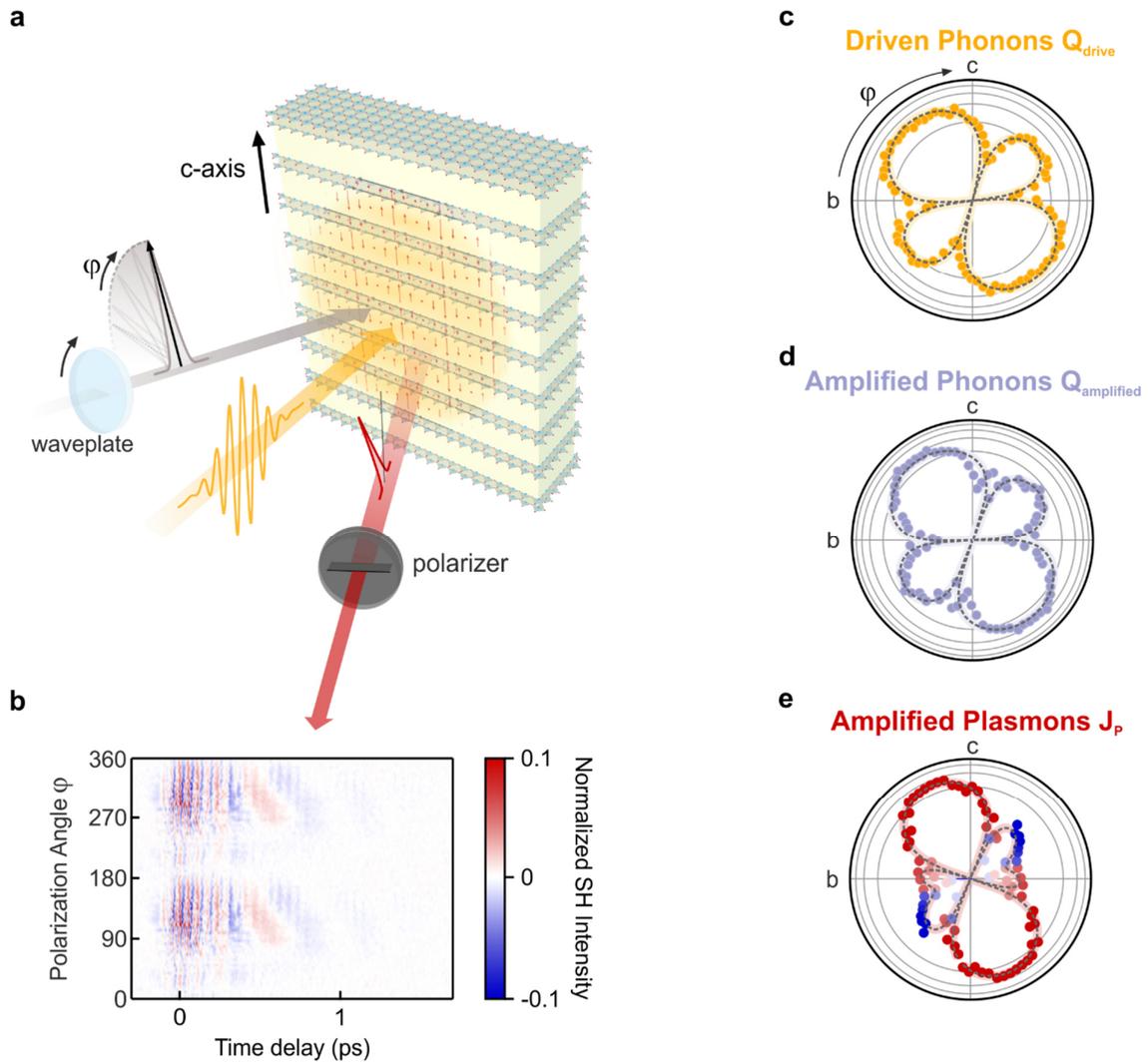

**Fig. S7 a,b** and **c**, Normalized polarimetry signal of the driven phonons (yellow dots), amplified phonons (grey dots) and amplified Josephson Plasmon Polariton (red and blue dots) for an analyzer oriented along the crystal *b*-axis, at one time-delay t = 500 fs. The polarimetry signal of the two sets of phonons can be reproduced by a fit to a *mm2* point group symmetry (dashed line) and the phase of the oscillations is polarization angle $\varphi$ independent. The polarimetry signal of the amplified Josephson plasmon agrees with a fit to point group *m* (dashed line). The phase of the polarimetry signal is indicated by the red and blue color-coding.

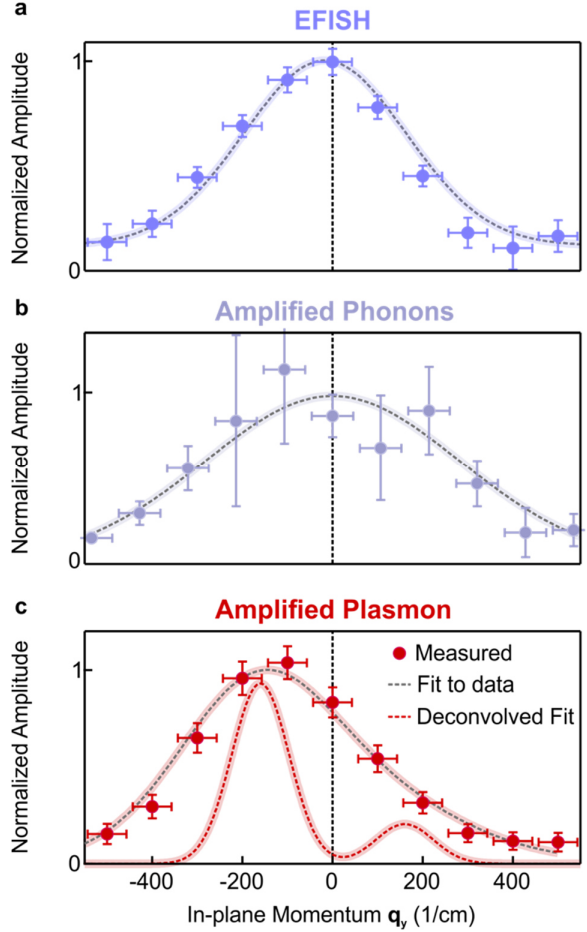

**Fig. S8 a** In-plane momentum distribution of the EFISH-component, **b** amplitude of the amplified phonon oscillations and **c** coherent Josephson Plasmon Polariton oscillations, as measured in the experiment sketched in Figure 4a of the main text. Data points are shown as blue, grey and red symbols. The Gaussian fit to the EFISH data, shown as a dashed grey line in **a**, reveals the divergence of the second harmonic beam. In panel **c**, the fit to the raw data and of the Josephson Plasmon Polariton amplitude and its $\omega_{IR} = \omega_{J1}(q_{plas}) + \omega_{J2}(-q_{plas})$ deconvolution are plotted as dashed grey and red lines, respectively. Error bars represent the standard deviation $\sigma$ of the amplitudes extracted by numerical fits. Horizontal error bars represent the standard deviation $\sigma$ due to the finite width of the measurement slit.

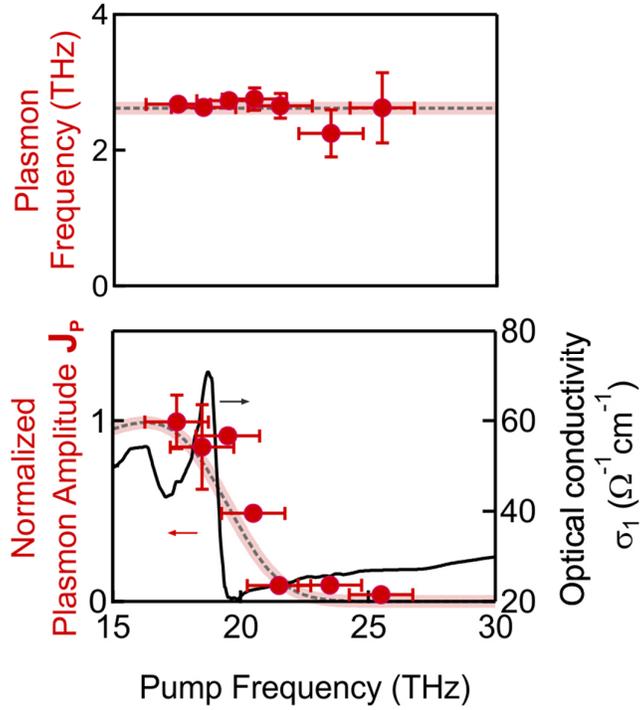

**Fig. S9** Amplitude and frequency of the low-frequency Josephson Plasmon Polariton in YBa$_2$Cu$_3$O$_{6.48}$ at different mid-infrared frequencies, shown as red points, for a fixed peak electric field of ~7 MV/cm. The amplitude increases towards the resonance of the apical oxygen infrared vibration. The real part of the optical conductivity is drawn as solid black line, and as a dashed line when convolved with the bandwidth of the excitation pulses. Error bars represent the standard deviation $\sigma$ of the JPP amplitudes extracted by numerical fits. Horizontal error bars represent the standard deviation $\sigma$ of the center wavelength determined by electro-optic sampling.

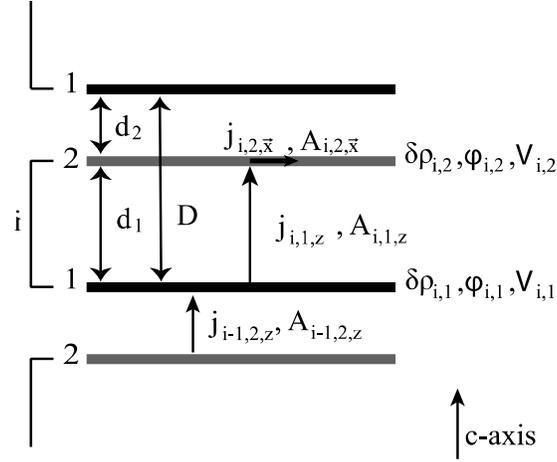

**Fig. S10** Schematic drawing of a bilayer superconductor. Variables $\delta\rho_{i,\lambda}$, $\phi_{i,\lambda}$, $j_{i,\lambda,\vec{x}}$, and $A_{i,\lambda,\vec{x}}$ are defined within layer $\lambda$ in unit cell $i$ and describe condensate density fluctuations, phase of the order parameter, parallel component of the superfluid current, electrostatic potential, and in-plane vector potential respectively. Variables $j_{i,\lambda,z}$, and $A_{i,\lambda,z}$ are defined between the layers and correspond to interlayer Josephson current and out of plane component of the vector potential, respectively.

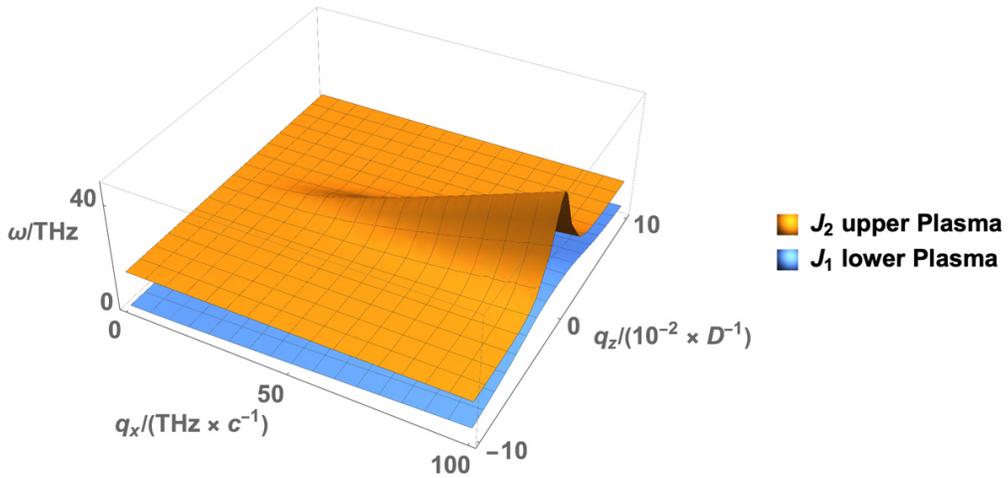

**Fig. S11:** Dispersion relation of the two lowest energy modes of equations (10)-(12) in the $\{q_x, q_z\}$-plane. At $q_z = 0$, the upper plasmon is strongly hybridized with the original photon mode. This results in the energy of the mode increasing rapidly along the $q_x$ axis with the slope approaching the speed of light. Away from $q_z = 0$ strong mixing with the photon is absent and the frequency of the mode decreases with increasing $q_x$.

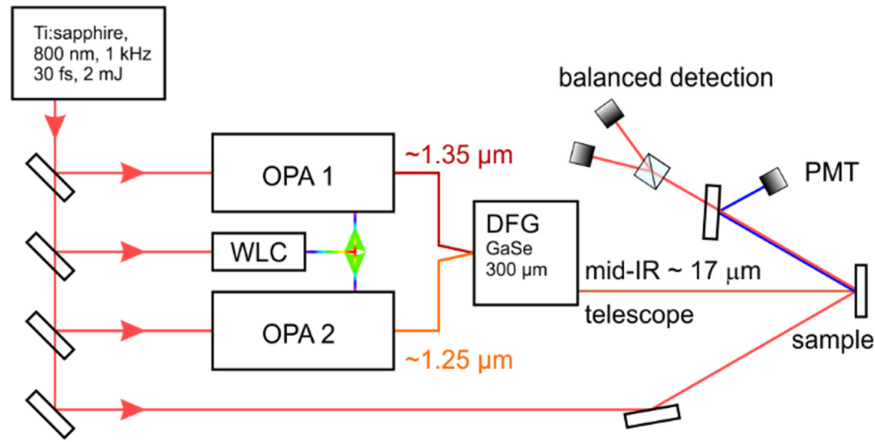

**Fig. S12:** Schematic drawing of the experimental setup. 30-fs pulses from a Ti:sapphire amplifier pump two optical parametric amplifiers (OPA), which are seeded by the same white light continuum (WLC) . Carrier envelope phase stable 3-µJ 150-fs pulses at 17 µm wavelength are generated by difference frequency mixing the two signal beams from the OPAs. These excitation pulses are focused onto the sample at spot size of ~65 µm and overlapped with the 800-nm probe pulses (35 µm spot size). Their time delay dependent second harmonic intensity and polarization rotation are detected by a photo multiplier tube and a balanced detection scheme, respectively.